\documentclass[a4paper,12pt]{article}
\usepackage{epsfig}

\begin{document}

\def\beq{\begin{equation}}
\def\eeq{\end{equation}}
\def\bea{\begin{eqnarray}}
\def\eea{\end{eqnarray}}

\def\dofourfigs#1#2#3#4#5{\centerline{
\epsfxsize=#1\epsfig{file=#2, width=7.5cm,height=7.0cm, angle=0}
\hspace{0cm}
\hfil
\epsfxsize=#1\epsfig{file=#3,  width=7.5cm, height=7.0cm, angle=0}}

\vspace{0.5cm}
\centerline{
\epsfxsize=#1\epsfig{file=#4, width=7.5cm,height=7.0cm, angle=0}
\hspace{0cm}
\hfil
\epsfxsize=#1\epsfig{file=#5,  width=7.5cm, height=7.0cm, angle=0}}
}

\def\dosixfigs#1#2#3#4#5#6#7{\centerline{
\epsfxsize=#1\epsfig{file=#2, width=6.5cm,height=5.5cm, angle=0}
\hspace{0cm}
\hfil
\epsfxsize=#1\epsfig{file=#3,  width=6.5cm, height=5.5cm, angle=0}}

\vspace{0.5cm}
\centerline{
\epsfxsize=#1\epsfig{file=#4, width=6.5cm,height=5.5cm, angle=0}
\hspace{0cm}
\hfil
\epsfxsize=#1\epsfig{file=#5,  width=6.5cm, height=5.5cm, angle=0}}

\vspace{0.5cm}
\centerline{
\epsfxsize=#1\epsfig{file=#6, width=6.5cm,height=6.cm, angle=0}
\hspace{0cm}
\hfil
\epsfxsize=#1\epsfig{file=#7, width=6.5cm, height=6.cm, angle=0}}
}

\def\dotwofigsa#1#2#3{\centerline{
\epsfxsize=#1\epsfig{file=#2, width=3cm,height=3.8cm, angle=0}
\hspace{0cm}
\hfil
\epsfxsize=#1\epsfig{file=#3,  width=8cm, height=7.5cm, angle=0}}
}

\def\dotwofigs#1#2#3{\centerline{
\epsfxsize=#1\epsfig{file=#2, width=6cm,height=8cm, angle=0}
\hspace{0cm}
\hfil
\epsfxsize=#1\epsfig{file=#3,  width=7cm, height=8cm, angle=0}}
}

\def\dofig#1#2{\centerline{
\epsfxsize=#1\epsfig{file=#2, width=12cm,height=9cm, angle=0}
\hspace{0cm}
}}

\def\dofigc#1#2{\centerline{
\epsfxsize=#1\epsfig{file=#2, width=15cm,height=8cm, angle=0}
\hspace{0cm}
}}

\def\dofigb#1#2{\centerline{
\epsfxsize=#1\epsfig{file=#2, width=15cm,height=8cm, angle=0}
\hspace{0cm}
}}

\def\dofiga#1#2{\centerline{
\epsfxsize=#1\epsfig{file=#2, width=4cm,height=4cm, angle=0}
\hspace{0cm}
}}

\newcommand{\dedouble}{ \stackrel{ \leftrightarrow }{ \partial } }
\newcommand{\deR}{ \stackrel{ \rightarrow }{ \partial } }
\newcommand{\deL}{ \stackrel{ \leftarrow }{ \partial } }
\newcommand{\ci}{{\cal I}}
\newcommand{\ca}{{\cal A}}
\newcommand{\Wp}{W^{\prime}}
\newcommand{\vep}{\varepsilon}
\newcommand{\kk}{{\bf k}}
\newcommand{\pp}{{\bf p}}
\newcommand{\hs}{{\hat s}}
\newcommand{\proj}{\frac{1}{2}\;(\eta_{\mu\alpha}\eta_{\nu\beta}
+  \eta_{\mu\beta}\eta_{\nu\alpha} - \eta_{\mu\nu}\eta_{\alpha\beta})}
\newcommand{\projm}{\frac{1}{2}\;(\eta_{\mu\alpha}\eta_{\nu\beta}
+  \eta_{\mu\beta}\eta_{\nu\alpha}) 
- \frac{1}{3}\;\eta_{\mu\nu}\eta_{\alpha\beta}}

%%%%%%%%%%%%%%%%%%%%%%%%%%%%%%%%%%%%%%%%%%%%%%%%%%%%%%%%%%%%%%%%%%%%%%%
\def\lsim{\raise0.3ex\hbox{$\;<$\kern-0.75em\raise-1.1ex\hbox{$\sim\;$}}} 

\def\gsim{\raise0.3ex\hbox{$\;>$\kern-0.75em\raise-1.1ex\hbox{$\sim\;$}}}

\def\Frac#1#2{\frac{\displaystyle{#1}}{\displaystyle{#2}}}
\def\no{\nonumber\\}
%-------------------------------------------------------------------------
\renewcommand{\thefootnote}{\fnsymbol{footnote}}

\rightline{CP3-07-03; \, \, FNT/T-2007/01; \, \, HIP-2006-01/TH}

{\Large
\begin{center}
{\bf Higgs Boson Production in Association with a Photon in 
Vector Boson Fusion at the LHC}
\end{center}}
\vspace{.3cm}

\begin{center}
Emidio Gabrielli$^{1}$, $\;$ Fabio Maltoni$^2$,
$\;$ Barbara Mele$^{3}$, $\;$ \\ Mauro Moretti$^{4}$, 
$\;$ Fulvio Piccinini$^{5}$, $\;$ and $\,$ Roberto Pittau$^{6}$\\
\vspace{.5cm}

$^1$\emph{Helsinki Institute of Physics,
     POB 64, University of Helsinki, FIN 00014, Finland
}
\\
$^2$\emph{Centre for Particle Physics and Phenomenology (CP3) \\
Universit\'{e} catholique de Louvain \\ 
Chemin du Cyclotron 2, 
B-1348 Louvain-la-Neuve, Belgium}
\\
$^3$\emph{INFN, Sezione di Roma,
and Dipartimento di Fisica, Universit\`a La Sapienza, \\
P.le A. Moro 2, I-00185 Rome, Italy}
\\
$^4$\emph{Dipartimento di Fisica Universit\`a di Ferrara, and INFN,  
Sezione di Ferrara, \\
via Saragat 1, I 44100, Ferrara, Italy}
\\
$^5$\emph{INFN, Sezione di Pavia, via A. Bassi 6, I 27100, Pavia, Italy}
\\
$^6$\emph{Dipartimento di Fisica Teorica, Universit\` a di Torino,
and INFN sezione di Torino, via P. Giuria 1, Torino, Italy}
\end{center}

\vspace{.3cm}
\hrule \vskip 0.3cm
\begin{center}
\small{\bf Abstract}\\[3mm]
\begin{minipage}[h]{14.0cm}
 Higgs boson production
 in association with two forward jets and a central photon at the CERN Large 
Hadron Collider is analyzed, for the Higgs boson decaying into a 
$b\bar{b}$ pair in the  $m_H \lsim 140$~GeV  mass region.
We study  both irreducible  and  main reducible backgrounds 
at parton level. 
Compared to the Higgs production via vector-boson fusion,
the request of a further photon at moderate rapidities
dramatically enhances the signal/background ratio. 
Inclusive cross sections for $p_{\rm T}^{\gamma}\gsim 20$~GeV 
can reach a few tens of fb's.
After a suitable choice of kinematical cuts, the cross-section ratio for 
signal and irreducible-background  can be enhanced up to $\gsim$~1/10, 
with a signal 
cross section of the order of a few  fb's, 
for  $m_H \sim120$~GeV. The request of a central photon radiation also enhances
the relative signal sensitivity to the $WWH$ coupling  with respect to the 
$ZZH$ coupling. 
Hence,
a determination  of the  cross section for the associated production of
a Higgs boson decaying into a $b\bar{b}$ pair plus a central photon in 
vector-boson fusion
could help in constraining  the  $b\bar{b}H$ coupling, and the
$WWH$ coupling as well.
 A preliminary study of QCD showering effects points to a further 
significant improvement of the signal  detectability over the background.
\end{minipage}
\end{center}
\vskip 0.3cm \hrule \vskip 0.5cm
\vskip 0.3cm
%%%%%%%%%%%%%%%%%%%%%%%%%%%%%%%%%%%
\section{Introduction}
%%%%%%%%%%%%%%%%%%%%%%%%%%%%%%%%%%%
Higgs boson search is one of the main tasks of
present and future collider experiments.
The Higgs mechanism, responsible for 
the electroweak symmetry breaking (EWSB) in the Standard Model (SM), 
predicts the 
existence of a scalar particle, the Higgs boson, that  is 
still eluding any direct experimental test.
The fact that the 
Higgs boson mass is not predicted in the  SM,  
ranging from the present experimental lower limit  of 
114 GeV~\cite{LEP}, up to the theoretical upper bound of about  $800$ 
GeV~\cite{hunters,Dashen:1983ts,Luscher:1988gc}, 
makes this search more difficult.
Despite the large theoretical uncertainty on its mass,  present
precision tests  of electroweak observables
indicate that the data consistency in the SM requires  
a light Higgs~\cite{LEP}, that is  in a range close to the present 
experimental lower limit. 
Moreover, in the minimal supersymmetric extension of the SM 
the lightest Higgs scalar is expected to be lighter than
about 135 GeV~\cite{upperbound}.
Clearly, the  discovery 
of the Higgs boson would be a fundamental handle in pinpointing 
the EWSB mechanism, possibly shedding some light on physics scenarios 
beyond the SM. 

Higgs boson production mechanisms at colliders
have been intensively analyzed in  the literature~\cite{Hrev}.
The fact that  Higgs couplings to particles
are proportional to their masses makes Higgs  
production mechanisms and search strategies quite peculiar. 
A light Higgs boson (with $m_H \lsim 135$ GeV) mainly decays 
into $b\bar{b}$ pairs, making its experimental 
search very challenging at hadron colliders, because of
  large  backgrounds from QCD $b$-quark production.
At the CERN Large Hadron Collider (LHC), 
the Higgs boson is expected to be produced with high rate 
via gluon or vector-boson fusion (VBF) mechanisms and associate 
$W(Z) H$ production. 
A possible way to tame the huge QCD background in the search of a 
light Higgs boson is to look at its rare decays. In particular,
the $H\to \gamma \gamma$~\cite{Haa},  $H\to \tau\tau$~\cite{Htautau}, 
and $H\to Z^\ast Z^\ast \to 4 \ell $ channels are particularly promising, 
allowing the observation of the signal in a moderate background environment.

On the other hand, it would be crucial 
to make at the LHC also a measurement of the $Hb\bar{b}$ 
coupling~\cite{Duhrssen:2004cv}. 
To this aim, Higgs production via VBF, with the Higgs boson decaying into a 
$b\bar{b}$ pair, has been analyzed in~\cite{higgsplb}. The VBF characteristic 
signature  is  the presence of two forward
jets with a typical  transverse momentum of the order of $ M_W$, 
allowing a quite effective reduction of the
background. Nevertheless,
different sources of QCD backgrounds  and hadronic effects, 
that are  hard to control, make the relevance of this channel 
for a  $Hb\bar{b}$  determination presently difficult to assess. 

Another potential channel  for a  $Hb\bar{b}$ coupling measurement is  
the associated production  $Ht\bar{t}$, where the Higgs boson is 
radiated by a top-quark pair~\cite{richter,drollinger}. 
Unfortunately, recent studies that include a more reliable 
QCD background estimate and dectector simulation,  
see for instance Ref.~\cite{CMS-TDR}, have
lowered the expectations on the discovery potential of this channel.

In this paper, we consider a further process that could help in 
determining the $H b\bar{b}$ coupling, that is the
Higgs boson  production
in association with a large transverse-momentum  photon 
(with $p_{\rm T}\gsim 20$ GeV) and two {\it forward} jets
\begin{equation}
pp\to H\, \gamma\, jj\to b \bar b \,\gamma \,jj\, + X\, ,
\label{process}
\end{equation}
with $H$ decaying to $b\bar{b}$, where at the 
parton level the final QCD parton 
is identified with the corresponding jet. 
The  Feynman diagrams  that are dominant for this process at the parton level 
are the ones involving VBF. They are shown in Figure~1, where the Higgs 
decay to $b\bar{b}$ is not included. In principle, final states 
$b \bar b\, \gamma\, jj\,$ can also arise from photon radiation 
from one of the two  $b$-quarks coming from the Higgs 
boson decay, via the process $pp\to H(\to b \bar b\,\gamma )\, jj$.
In our study,
we will not include the latter set of diagrams, since the requirement of 
a large $p_{\rm T}$ photon would shift in that case the $b\bar b$ invariant 
mass outside the experimental $b\bar b$ mass resolution window around 
the Higgs mass. We will then assume that the effect of diagrams with 
photons arising from the final $b$'s  will vanish after
applying  the Higgs mass constraint in the event analysis of the 
$b \bar b\, \gamma\, jj\,$ final state.
\begin{figure}[tpb]
\begin{center}
\dofig{3.1in}{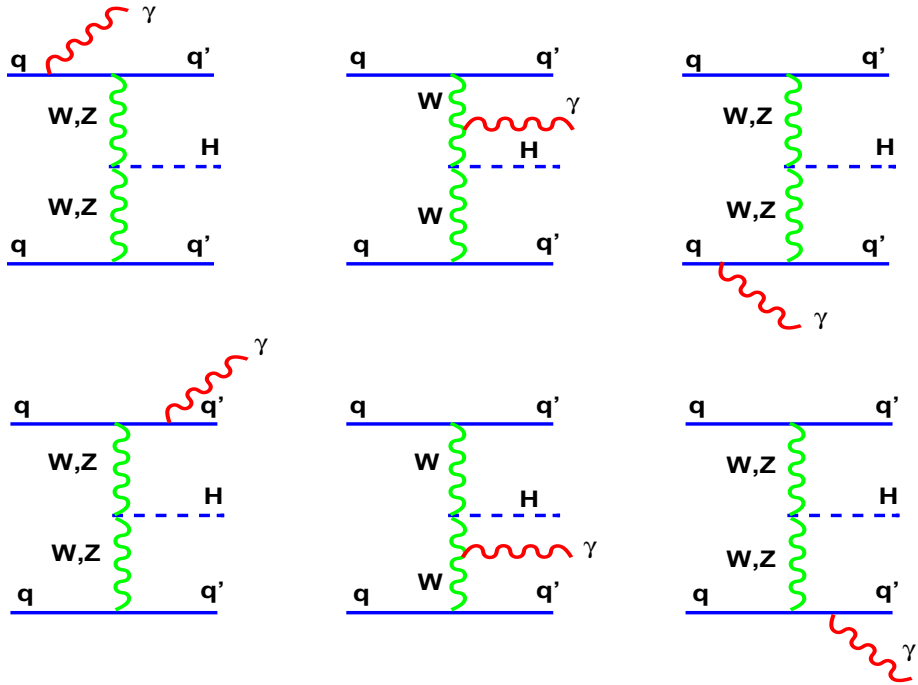}
\end{center}
\caption{\small Tree-level $t$-channel Feynman diagrams for  Higgs boson 
production in the process $pp\to H\,\gamma\, jj$. 
Here, $q$ and $q^{\prime}$ stand for different
light quarks ($u,d,s,c$), while  $q=q^{\prime}$ when a Z boson is exchanged.}
\end{figure}

There is a number of advantages in considering this QED higher-order 
variant of the VBF Higgs production process $pp\to H(\to b\bar b) \; jj$.
The fact that the production rate is penalized by the electromagnetic
coupling is compensated by a few peculiarities of the channel 
in Eq.~(\ref{process}).

First of all, the presence of an additional  high $p_{\rm T}$ photon can 
improve the triggering efficiencies for multi-jet final 
states, such as those needed to select $pp\to H(\to b\bar b) \; jj$  events. 
Second, there is a large gluonic component entering the partonic 
processes giving rise to the QCD backgrounds to the 
$b \bar b \,\gamma \,j j$ final state. As a consequence,
the QCD backgrounds are in general much less  {\em active}
in  radiating  a large $p_{\rm T}$ photon with respect to the VBF signal. 
In addition there are further dynamical effects that 
dramatically suppress the radiation of a central photon in the 
irreducible QCD background to  $b \bar b \,\gamma \,jj$ with 
respect to the VBF channel, as we shall see in Section~3.

The requirement of a central photon is also expected to 
strongly reduce the contribution arising from alternative 
Higgs boson production processes, such as the one
coming from the virtual gluon fusion 
$g^{\ast} g^{\ast} \to H$ diagrams~\cite{delduca}, 
with a  photon radiated from any external 
quark leg\footnote{Note that  the  virtual gluon fusion 
diagrams $g^{\ast} g^{\ast} \to H \gamma$,
where a photon is attached to the top-quark loop, 
exactly vanishes according to charge conjugation invariance.}. 
This expectation is confirmed in our analysis.

The requirement of having two jets with high $p_{\rm T}$ in the 
final states also suppresses any  contamination
from the one-loop $q\bar{q} \to H\gamma$ process, that has 
anyhow a quite low cross section at the LHC~\cite{Ha}.

There is a further issue that increases the potential of the process in 
Eq.~(\ref{process}) for the determination of Higgs boson couplings.
We will see that the requirement of a central photon depletes the
$HZZ$ amplitudes with the respect to the $HWW$ ones in Figure~1.
As a consequence,  the relative sensitivity to the $HWW$ coupling 
is considerably  increased in the radiative channel. Hence, a measurement
of the $b \bar b \,\gamma \,jj$ rate could lead to a combined 
determination of the Higgs boson couplings to  $b$ quarks and $W$
vector bosons, with less contamination from the $HZZ$ coupling 
uncertainties.

In~\cite{rainwater},
another method  for  extracting information on the $H b\bar{b}$ coupling  
has been suggested through the Higgs boson
production in association with a $W$ plus two forward jets.
In this case, the main reducible  backgrounds are given by the
$W b\bar{b} j j $ and $t\bar{t} j j $ final states, 
and can be suppressed 
by proper kinematical cuts. 
However, the requirement of a leptonic decay of the $W$ limits 
the event statistics. In the following, we will compare
the corresponding rates with the $b \bar b \,\gamma \,jj$ 
ones that we analyze here.

The plan of the paper is the following. 
In Section~2, we go through the main kinematical and 
dynamical characteristics of the process in Eq.~(\ref{process}). 
We also discuss the features of the main QCD irreducible background. 
Section~3 is devoted to the discussion of the destructive 
quantum interference mechanism that is largely responsible  
both for the improvement 
in the signal-to-background ratio of the channel considered, 
and for its increased sensitivity to the $WWH$ coupling. 
In Section~4, the signal rates are computed at parton level 
for a set of kinematical cuts 
that optimizes the signal/background ratio, restricting the 
analysis to the case of the irreducible background. Some relevant 
kinematical distributions are also shown, and compared with the ones 
for the basic VBF process $H j j$. In Section~5, the main reducible 
background channels are included in the analysis. 
Some preliminary study of parton-shower effects and jet veto strategies, 
that turn out to improve the signal detectability, is performed in 
Section~6. Finally, in Section~7, we draw our conclusions.

%%%%%%%%%%%%%%%%%%%%%%%%%%%%%%%%%%%%%%%%%%%%%%%%%%%%%%%%%%%%%%%%%%%%
\section{Signal and irreducible background}
\label{sect2}
%%%%%%%%%%%%%%%%%%%%%
\begin{table}
\begin{center}
\begin{tabular}{||l|l|l|l|l||}\hline 
 $m_H $~(GeV)  & 110 &  120 & 130 & 140
\\  \hline
$\sigma(H
\gamma jj) \; \; [fb]$  & 67.4  & 64.0   & 60.4  &  56.1
 \\ \hline
${\cal BR}(H\to b \bar b)$ & 0.770 & 0.678 & 0.525 & 0.341 \\ \hline
\end{tabular}            
\caption{\label{inclusive} Cross sections for the
$H\,\gamma\, jj$ signal at LHC, for $p_{\rm T}^\gamma \geq 20$~GeV, 
$\Delta R_{\gamma j}> 0.4$, and  a cut $m_{jj} > 100\, {\rm GeV}$ on 
the invariant mass of the final quark pair. Also shown are the 
Higgs boson branching ratios to $b\bar b$
(computed through HDECAY~\cite{hdecay}), that are not included in 
the cross sections shown.}
\end{center}
\end{table}
Cross sections for the $H\, \gamma\, jj$ production
at $\sqrt S =14$~TeV
are shown in Table~\ref{inclusive}.  In order to
present results as inclusive as possible only a minimal set of
kinematical cuts is applied
($\Delta R_{\gamma j}> 0.4$,
$p_{\rm T}^\gamma \geq 20$~GeV, and 
$m_{jj} > 100\, {\rm GeV}$).  The cut on the invariant mass of the final
quark pair ($m_{jj}$) avoids the contribution from {\it resonant}
$HW\gamma, \,HZ\gamma$ associated production.
The Higgs boson branching ratios to $b\bar b$, which are
not included in the cross section results,
(computed through HDECAY~\cite{hdecay}), are also shown.
The full tree-level matrix elements for the electroweak process
$pp\to H\, \gamma\, jj\,$
%in Figure~1
have been computed independently with ALPGEN~\cite{alpgen},
and MadEvent~\cite{madevent}. Details on the values of the input
parameters, such PDF's and scales are given in Section~4.

For comparison, the inclusive $H\, W\, jj$ cross section, with a
further cut
$\vert m^2_{ik} \vert > 100\, {\rm GeV}^2$ on the invariant mass of  
any $ik$
initial-final quark pair to  avoid singularities due to $t$-channel
virtual photons,
is  73~fb, for
$m_H = 120$ GeV. Requiring  the leptonic (either  $\mu^\pm$ and
$e^\pm$) signature for the
$W$~\cite{rainwater}, one ends up with a rate for the
$H\, \gamma\, jj$
signal of about 4 times the
$H\, \ell\nu\, jj$ rate, in the relevant  $m_H$  range.
This factor increases, when more realistic kinematical
cuts are applied (see Section~4).

Since the LHC detector capabilities are not completely
settled yet, in what follows we will assume two
different setups for the high $p_{\rm T}$ photon  threshold: either
$p_{\rm T}^\gamma \geq 20$~GeV or $p_{\rm T}^\gamma \geq 30$~GeV, 
the latter being more conservative in the case the photon identification 
and/or electromagnetic trigger requires a larger photon transverse momentum.

It is useful to  recall here the main kinematical properties 
of a typical VBF event, that is $pp\to H\, jj$, and 
the corresponding backgrounds, 
assuming the $H\to b\bar{b}$ decay, for $m_H\lsim 140$ GeV.
\noindent
We will see in the next section that the request of a further 
large $p_{\rm T}$ photon tends to enhance the peculiar kinematical 
features of a VBF event.
\begin{figure}[tpb]
\begin{center}
\dofiga{3.1in}{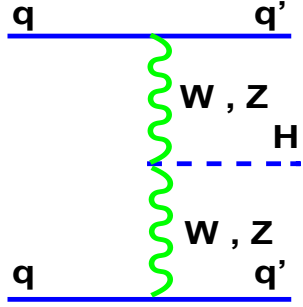}
\end{center}
\caption{\small 
Feynman diagram at parton level for the Higgs boson 
production via VBF in the process $pp\to H \,jj$.
Here, $q$ and $q^{\prime}$ stand for 
 light quarks  $(u,d,s,c)$, while  $q=q^{\prime}$ when a Z boson is exchanged.}
\end{figure}
In Figure~2, the basic partonic process  for a VBF
event, namely $q q \to q q H$, is shown, 
where the two final quarks  hadronize in two  jets.
The typical VBF signature 
consists of two jets with large invariant mass, widely 
separated in rapidity, and with a typical transverse
momentum of $p_{\rm T}\sim 40$ GeV, the Higgs boson decay products 
lying at intermediate rapidities. In particular,
while one of the two jets is produced  quite along the beams, 
with a pseudorapidity ($\eta$) distribution peaked around $\eta\sim 3-4$,
the other one is mainly backward, although still in the central detector,
with  $|\eta|\lsim 2.5$. 
\begin{figure}[tpb]
\begin{center}
\dofigc{3.1in}{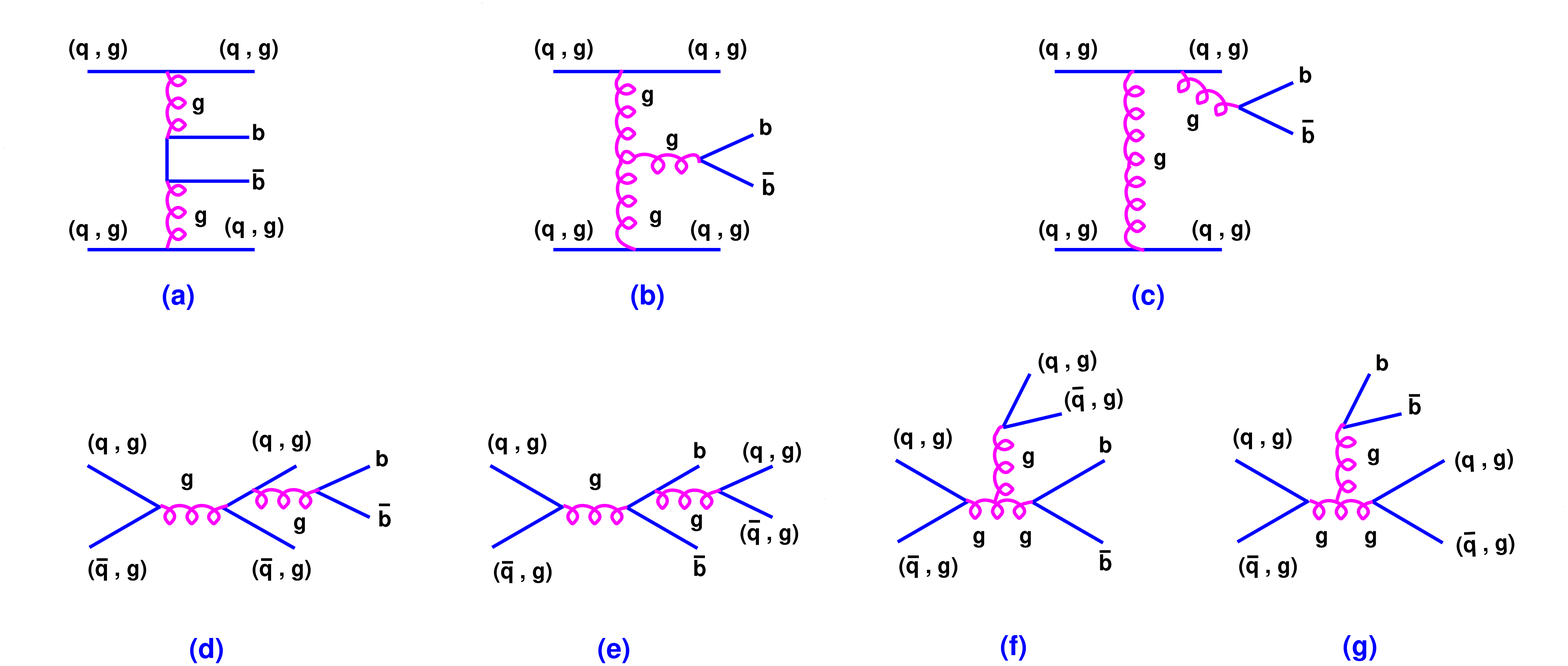}
\end{center}
\caption{\small 
Representative classes of Feynman diagrams 
contributing, at parton level, to the background process
$pp\to b\bar{b} \,jj$. Here, $q$ and $g$ stand for a
light quark $(u,d,s,c)$ and gluons respectively. 
The virtual gluon connecting
the $b\bar{b}$ pair in $c)-d)$ or the $(q,\bar{q})$ and $(g,g)$ 
pairs in $e)$, is understood to be attached in all possible ways 
to the initial and final parton. Crossed diagrams are not shown. }
\end{figure}

For the Higgs boson decaying to a $b\bar{b}$ pair, the main 
background to the basic VBF process comes from the QCD
production of  the final state $b\bar{b} jj$,  whenever the 
$b\bar{b} jj$ kinematical characteristics approach the typical 
VBF configuration. Here,  $j$ stands for a jet originating from 
either a light quark
$(u,d,s,c)$ or a gluon.
Seven representative classes for the $b\bar{b} jj$ background
Feynman diagrams at parton level  are  
given in Figure~3 (a$-$g), where all external partons, 
but the  $b\bar{b}$ final pair, 
can be either quarks or gluons.

Although the inclusive cross section for the 
$pp\to H(\to b {\bar b})\, jj$ signal is quite large, of the
order of a  few pb's, the extraction of
the signal from the background is not at all straightforward, 
being the latter dominant over the signal 
by a few orders of magnitude.
However, with a suitable choice of kinematical cuts,
the ratio of the expected signal event number $(S)$ over the 
background ones $(B)$ can be substantially enhanced.
By imposing a large invariant mass cut for the two-forward-jet 
system [i.e., $m_{jj}\gsim 
{\cal O}(1)$ TeV], a minimal $p_{\rm T}^{j}$ of a few tens GeV's, and 
requiring  the $b \bar b$ invariant mass to be  around $m_H$
within the $m_{b \bar b}$ experimental resolution, one can obtain
a signal significance ($S/\sqrt{B}$) 
of the order of $S/\sqrt{B}\sim 3-5$, assuming an integrated 
luminosity of 600 fb$^{-1}$~\cite{higgsplb}.

Let us now consider the  VBF Higgs production  when  a further 
central photon is emitted, namely $pp\to H\, \gamma\, jj$. 
The Feynman diagrams for the signal  in 
Figure~1 are now to be confronted with the QCD background
corresponding to the requirement of a further high $p_{\rm T}$ photon
in the diagrams in Figure~3. For instance, in Figure~4 (a$-$g), 
we show the relevant diagrams  for one of the  leading
classes of contributions, i.e. the one related to  Figure~3 (a). 
\begin{figure}[tpb]
\begin{center}
\dofigb{3.1in}{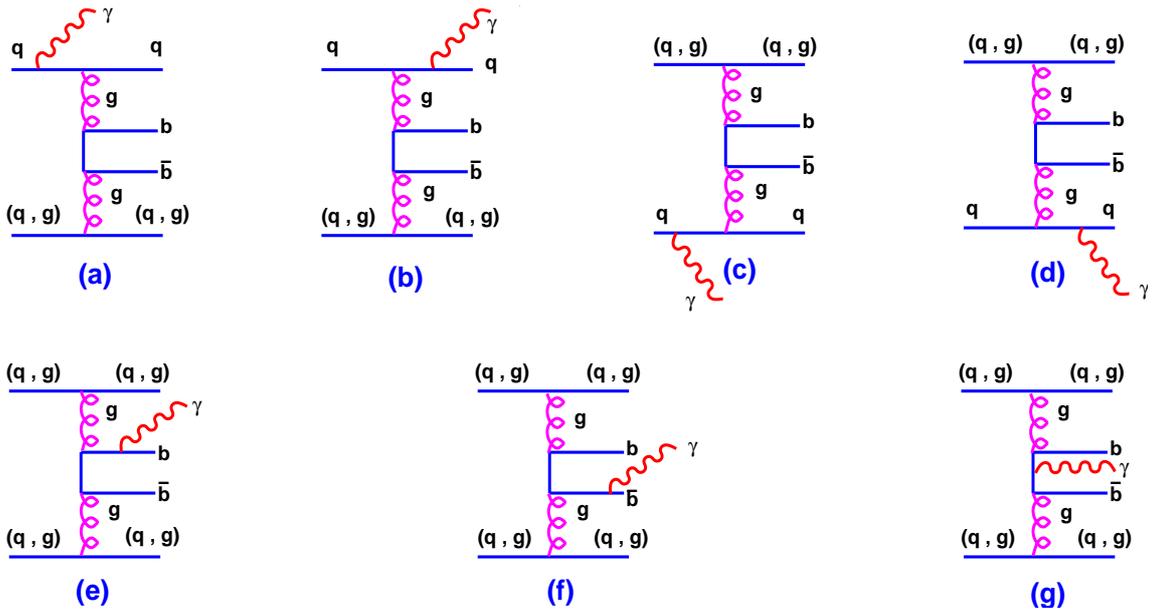}
\end{center}
\caption{\small Representative class of Feynman diagrams
at parton level for the 
the background process $pp\to b\bar{b} jj+\gamma$, corresponding to 
the photon emission for the class of diagrams in Figure~3 (a).
Here q and g stands for a
generical light quark $(u,d,s,c)$ and gluon respectively.}
\end{figure}

According to the usual pattern of QED corrections,
one might expect the request of a further hard photon to keep
the  relative weight of signal and background quite stable. 
Were this the case, the rates for $pp\to H\, \gamma\, jj$ and
its background would be related to a ${\cal O}(\alpha)$ rescaling
of the rates for the $H\, jj$ signal  and its background, respectively,
where $\alpha$ is the fine electromagnetic structure constant .
Then, the $S/B$ ratio would not be much affected.
On the other hand, both  the  $H\, \gamma\, jj$ signal and its 
background statistics would 
 decrease according to the  rescaling factor ${\cal O}(\alpha)$.
Consequently, if $(S/\sqrt{B})|_{H(\gamma)\,jj}$
is the signal significance for the VBF process
(with) without a central photon,
one would expect  the signal significance for  
$pp\to H\, \gamma\, jj$ to fall down as  
$(S/\sqrt{B})|_{H\gamma \,jj} \sim \sqrt{\alpha}\, 
(S/\sqrt{B})|_{H\,jj}\lsim 1/10\,(S/\sqrt{B})|_{H\,jj} $ 
with respect to the basic VBF process. 
On this basis,  one would conclude that 
there is no advantage in considering the $H\, \gamma \, jj$
variant of the  $H\, jj$ process, apart from the fact that the 
presence of a hard photon
in the final state can improve the triggering  efficiency of 
the detectors.

In the next section, we will show that this pattern does not 
hold in general.
The QED naive expectations definitely hold  for inclusive processes,
but they do not necessarily apply when restricted regions 
of phase space are considered.
Indeed, we will see that the naive QED  rescaling fails 
for the main background processes considered here, when  
relevant sets of kinematical cuts  are imposed.
In particular, the requirement of a further central 
 photon gives rise to a dramatic increase  (by more than 
one order of magnitude)
in the $S/B$ ratio, 
while the signal cross section roughly follows the naive 
QED rescaling.
%%%%%%%%%%%%%%%%%%%%%%%%%
\section{Destructive interferences in central photon emission}
\label{sect3}
%%%%%%%%%%%%%%%%%%%%%
We will now go through the main {\it partonic} components 
of the irreducible
QCD $\,b\bar{b} jj$ background  to the VBF process, in order 
to study how the request of a further central photon affects 
each of them, and the overall balance among them.

In our study of the irreducible background,  we will not consider the
electroweak production of 
$b \bar{b} jj$ final states (nor its extension to the 
$b \bar{b} \gamma\, jj$ case), through, e.g., the mediation of 
a $Z^{({\ast})}/\gamma^\ast \to b \bar{b}$ decay. 
The latter contributions  have cross sections not much larger than  
those of the signal considered here (see Section~4, and 
\cite{higgsplb}). 
On the other hand, their typical $b \bar b$ 
invariant mass is well below the Higgs boson mass range we are considering in 
the present study. Hence, we do not expect any contamination from 
this background for the foreseen experimental resolution on $m_{b \bar b}$. 

In the first two columns of Table~2, we show the absolute 
($\sigma_i$) and  fractional  ($\sigma_i/\sigma$)
  cross sections for the  QCD $\,b\bar{b} jj$
background corresponding to the different classes of partonic 
processes.
 An {\it optimized} set of kinematical 
cuts enhancing  the $S/B$ ratio
[that will be discussed subsequently in Section~4, cf. 
Eqs.~(\ref{eq:basic}) and (\ref{eq:optimized}), set~1] is applied.

\noindent
The main effective constraints for the  $b\bar{b} jj$ channel, 
apart from the jet isolation, are
a large invariant mass for the final 
$jj$ system, namely $m_{jj}\gsim 800$~GeV, 
and the restriction on  the  $b\bar{b}$ 
invariant mass to be inside a  window of 
 $ m_H(1\pm10\%)$.  After applying these cuts, and considering 
the $m_H=120$ GeV case, the total QCD $b\bar{b} jj$ cross section
 turns out to be $\sigma \simeq 103$ pb.
\begin{table}
\begin{center}
\begin{tabular}{||l||l|l||l|l||}\hline
sub-processes   & $\sigma_i\;\;({\rm pb})$ & $\sigma_i/\sigma$ & 
 $\sigma_i^{\gamma}\;\;({\rm fb})$
& $\sigma^{\gamma}_i/\sigma^{\gamma}$ \\  \hline
$g q\to b\bar{b}\, g q\, (\gamma)$ & 57.2(1)  & 55.3 
\% &  17.3(1) & 51.6 \% 
\\ \hline
$g g\to b\bar{b}\, g g\, (\gamma) $  & 25.2(1) & 24.4 
\% &   3.93(3) & 11.7 \% 
\\ \hline
$q q^{\prime}\to b\bar{b}\, q q^{\prime}\, (\gamma) $  
& 7.76(3) &7.5 \%  
& 4.04(2) & 12.1 \% \\ \hline
$q q\to b\bar{b}\, q q\, (\gamma) $ & 6.52(2) &6.3 \%  
&     4.49(3) & 13.4 \%
\\ \hline
$q \bar{q}^{\prime}\to b\bar{b}\, q \bar{q}^{\prime}\, (\gamma) $  
& 4.60(2) &4.4 \%  
&  2.28(2) & 6.8 \% \\ \hline
$q \bar{q}\to b\bar{b}\, q \bar{q}\, (\gamma) $  
& 2.13(2) & 2.1 \%  
& 1.21(2) & 3.6 \% \\ \hline
$gg\to b\bar{b}\, q \bar{q}\, (\gamma) $  & 0.0332(7) & 0.03 \% 
&  0.124(3) & 0.37 \% \\ \hline
$q \bar{q}\to b\bar{b}\, gg\, (\gamma) $  & 0.0137(2) & 0.01 \%  
& 0.094(2) & 0.28 \% \\ \hline
$q \bar{q}\to b\bar{b}\, q^{\prime} \bar{q}^{\prime}\, (\gamma) $  
& 0.000080(3) & 0.00007 \%  &    0.00080(8) & 0.002 \%
\\ \hline
\end{tabular}            
\caption{\label{partial} Partial contributions  
$\sigma_i$ ($\sigma_i^{\gamma}$) in pb (fb)
of the partonic sub-processes to the total cross section
 $\sigma=103$ pb ($\sigma^{\gamma}=33.5$ fb), corresponding 
to the background process
$pp\to b\bar{b}\, jj \, (\gamma)$, for $m_H=120$ GeV.
Optimized kinematical cuts (set 1), as defined in Section~4, 
Eqs.~(\ref{eq:basic}) and (\ref{eq:optimized}),
are implemented. The numbers in parenthesis
correspond to the numerical errors on the last digit.
}
\end{center}
\end{table}
%%%%%%%%%%%%%%%%
\begin{table}
\begin{center}
%%%%%%%%%%%%%%%%
\begin{tabular}{||l||l|l||}\hline
sub-processes    & 
 $\sigma_i^{\gamma}[{\rm no}\; b\; {\rm rad}]\;\;({\rm fb})$
& $\sigma^{\gamma}_i[{\rm no}\; b\; 
{\rm rad}]/\sigma^{\gamma}[{\rm no}\; b\; {\rm rad}]$ \\  \hline
$g q\to b\bar{b}\, g q \gamma$ 
           &  8.19(6)  & 47.8 \% 
\\ \hline
$g g\to b\bar{b}\, g g \gamma$  
               &   0 & 0 \% 
\\ \hline
$q q^{\prime}\to b\bar{b}\, q q^{\prime} \gamma$  
                 &   2.80(2) & 16.4 \% 
\\ \hline
$q q\to b\bar{b}\, q q \gamma$ 
                 &   3.49(3) & 20.4 \%
\\ \hline
$q \bar{q}^{\prime}\to b\bar{b}\, q \bar{q}^{\prime}\gamma $  
                  &   1.57(2)& 9.2 \% 
\\ \hline
$q \bar{q}\to b\bar{b}\, q \bar{q} \gamma$  
                 &   0.87(1)& 5.1 \% 
\\ \hline
$gg\to b\bar{b}\, q \bar{q} \gamma$  
          &   0.10(2) & 0.6\% 
\\ \hline
$q \bar{q}\to b\bar{b}\, gg \gamma$  
         &   0.096(2) & 0.6 \% 
\\ \hline
$q \bar{q}\to b\bar{b}\, q^{\prime} \bar{q}^{\prime} \gamma$  
          &   0.0009(1) & 0.005 \%
\\ \hline
\end{tabular}            
\caption{\label{partial1} Partial contributions  
$\sigma_i^{\gamma}[{\rm no}\; b\; {\rm rad}]$~(fb)
of the partonic sub-processes to the total cross section
 $\sigma^{\gamma}[{\rm no}\; b\; {\rm rad}]$ =17.12(6)~fb for 
to the background process
$pp\to b\bar{b}\, jj \, \gamma$, when the photon radiation 
off $b-$quarks 
is switched off. The general setup is the same as in Table~2.}
\end{center}
\end{table}
Most of this cross section is due 
to the classes of diagrams involving  gluons in the
$t$-channel, as represented in Figure~3 (a$-$c). 
In particular, the latter give the main contributions to the cross 
sections for
the subprocesses $gg\to b\bar{b}+ gg$ and $gq\to b\bar{b}+ gq$,
$qq\to b\bar{b}+ qq$, and $qq^{\prime}\to b\bar{b}+ qq^{\prime}$, 
where $q^{\prime}\neq q$. 
Subleading QCD
contributions in Figure~3 (d$-$g) come from the fusion of initial
partons in $s$-channel type diagrams, like in  $gg\to b\bar{b}+ q\bar{q}$, 
$q\bar{q}\to b\bar{b}+ gg$, 
$q\bar{q}\to b\bar{b}+ q^{\prime}\bar{q}^{\prime}$,
with $q^{\prime}\neq q$. 
Indeed, the $s$-channel propagator
depletes  these contributions with respect to diagrams (a$-$c), 
when a large invariant mass for the $jj$ system is required.

Then, in the third and fourth columns of
Table~2,  we report
the absolute ($\sigma^\gamma_i$) and  fractional 
($\sigma^\gamma_i/\sigma^\gamma$) production rates, respectively, 
for the different classes of diagrams contributing to the  QCD 
$b\bar{b} \,\gamma \,jj$
background to the $H\,\gamma\, jj$ signal. The 
{\it optimized} set of kinematical 
cuts (as in Section 4,  Eqs.~(\ref{eq:basic}) 
and (\ref{eq:optimized}), set 1) is applied
for  $m_H=120$~GeV.
The corresponding  total QCD $b\bar{b}\, \gamma\,  jj$ cross section 
is $\sigma^{\gamma}\simeq 33$~fb.

One can see that the leading contribution to  total cross sections
is provided by the subprocess $g q\to b\bar{b}\, g q (\gamma)$, giving
more than 50\% in both the $b\bar{b}  \,jj$ and $b\bar{b} \,\gamma \,jj$
cases.
Next comes  the $gg\to b\bar{b} gg  (\gamma)$ channel,
with contributions of roughly 24\% and 12\%  to the 
$\sigma$ and $\sigma^\gamma$
total cross sections, respectively.
The set of kinematical cuts  imposed
(in particular the large invariant mass of the $jj$ system)
requires quite large partonic energy fractions for the initial partons.
This should suppress the relevance of gluon initiated channels with 
respect to the valence-quark initiated channel. Nevertheless, 
gluon initiated processes keep a leading role in contributing 
to the total background cross sections.

Now we can see why requiring a further central 
photon in VBF dramatically improves the detectability of 
the signal over the background.
On the one hand,  the requirement of a $p_{\rm T}^\gamma \gsim 20$ GeV 
central photon in the Higgs boson VBF production decreases the production 
rates by a bit less than two  orders of magnitude ({\sl i.e.} accordingly 
to the naive QED scaling expectations),
as we will see in  Section~4 (cf. Tables~4 and 5).
On the other hand, Table~2 shows that all the classes of 
partonic subprocesses
that contribute by more than 1\% to the irreducible-background 
cross sections
are reduced by more than three orders of magnitude as an effect of 
requiring a further  large $p_{\rm T}$
photon, giving an overall reduction factor in the background 
cross section
of $\sigma^\gamma/\sigma\sim1/3000$.

\noindent
In particular, we can see that the cross section for the  
second leading source of 
background for the basic VBF, that is the $gg\to b\bar{b} gg$ channel, is 
reduced by more than a factor 1/6000 after the photon 
requirement, falling from
$\sim 25$~pb down to $\sim 4$~fb for the 
$gg\to b\bar{b} gg\,\gamma$ channel. 
This is because gluons are neutral, and 
photons can only be emitted by the $b\bar{b}$ pair in this channel. 
Furthermore, the photon  radiation by $b$ quarks 
is also naturally suppressed by the down-quark electric charge.

The contribution arising from  photon emission in the 
subprocesses $gq\to b\bar{b}+ gq$,
$qq^{\prime}\to b\bar{b}+ qq^{\prime}$,  and $qq\to b\bar{b}+ qq$
might  in principle be larger. Nevertheless, as shown in Table~2, 
it is reduced approximately by a factor 1/3300, 1/2000, and 1/1500
for the $gq$,  $qq^{\prime}$ and $qq$ initial states, respectively, which 
is still an order of magnitude below the 
naive expectation for QED radiative  corrections.

This result can be explained in the following way.
When the photon is forced to be emitted in the central 
rapidity region (here, we are 
requiring in particular for the photon pseudorapidity 
$|\eta_\gamma| \leq 2.5$), 
a destructive quantum interference arises between 
the photon emission 
off the initial quark radiating a gluon 
(or any other neutral vector boson) in the $t$ channel, 
and the photon emission off the corresponding final quark. 
For instance,  in the set of  seven diagrams  shown in Figure~4,
the three subsets of diagrams  [(a) + (b)],  [(c) + (d)]
and [(e)+(f)+(g)] are separately  gauge invariant.
One can check (for example in the eikonal approximation) that, inside
the first two subsets, there is a strong destructive interference
 for   photons emitted outside the typical 
radiation cone around the initial/final partons
between  the diagrams  (a) $-$  (b) and  (c) $-$ (d) 
in Figure~4. Since in $t$-channel diagrams final partons are mostly 
approximately collinear
to the initial ones, requiring central $|\eta_\gamma|$'s 
indeed pushes photons
outside the typical radiation cone. 

\noindent
In the third subset of diagrams, [(e)+(f)+(g)] in Figure~4, 
the photon is emitted from a $b$-quark line. For the 
$b$-quark emission, 
no destructive interference is expected, and its contribution 
turns out to be important, 
and even dominant, in many classes of diagrams in Table~2, even if it
is controlled by the down-quark electric charge. This is shown in 
Table ~3,
where we compute again the entries of Table~2 for the radiative 
background 
$b\bar{b} \,\gamma \,jj$   after switching  off the photon radiation 
from  $b-$quarks
in the complete matrix element.
 The background total cross section falls  down to
 $\sigma^{\gamma}[{\rm no}\; b\; {\rm rad}]$ =17.3(3)~ fb in this case.
\begin{figure}[tbp]
\begin{center}
\epsfig{file=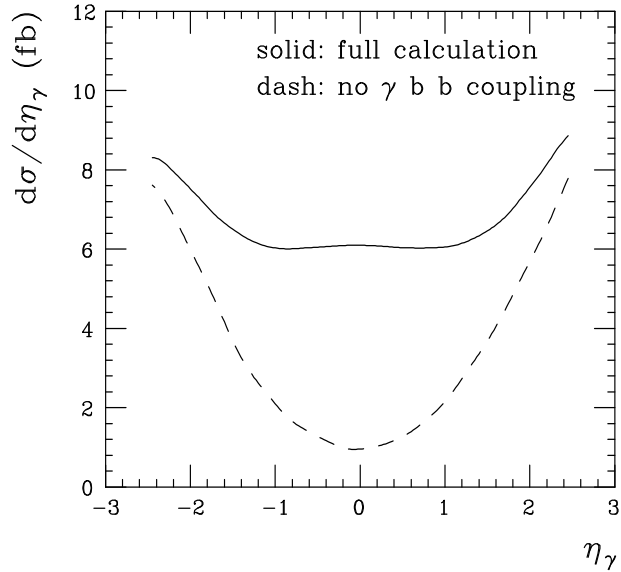,width=8cm,height=7.5cm, angle=0}
\end{center}
\caption{\small \label{detagamma} 
Effect of switching off the photon radiation from $b$ quarks
in the irreducible QCD background $pp\to b\bar{b} \gamma\, jj$
on the photon pseudorapidity distribution. 
The plot refers to the case $m_H=120$~GeV. Solid line: full calculation
for the {\it optimized} kinematical cuts (set 1) defined in Section~4, 
Eqs.~(\ref{eq:basic}) and (\ref{eq:optimized}).
Dashed line: same as before, assuming vanishing $b-$quark coupling to photons.}
\end{figure}

The destructive interference effect for central hard photons is 
even more  manifest
in Figure~5, where we plot the photon pseudorapidity distribution for 
the irreducible QCD background $pp\to b\bar{b} \gamma\, jj$ (solid curve).
The effect of switching off the photon radiation from $b$ quarks 
in the complete
matrix element of the process is shown by the dashed line.
The {\it optimized} kinematical cuts (set 1) defined in Section~4 
[Eqs.~(\ref{eq:basic}) and (\ref{eq:optimized})]  and $m_H=120$~GeV 
are assumed.
From the plot, one can  see the relative weight of the photon 
radiation off  
initial/final (non $b$) quarks. The dashed line clearly shows a 
deep corresponding to the photon central rapidity region, and 
arising from the destructive interference between initial and 
final quarks radiation. This deep is only partially filled by 
the photon radiation off $b$ quarks. We also checked that the 
same suppression at 
$\eta_\gamma\sim 0$ is observed for lower  $p_{\rm T}^\gamma$ 
values, by relaxing the $p_{\rm T}^{\gamma, cut}$ down to values 
as low as 5 GeV.
In fact, the behaviour of a  photon with a $p_{\rm T}$ as low as  
20~GeV at LHC energies is already 
well described by the eikonal approximation.

For the signal case of the $H\,\gamma\,jj$ production 
(Figure~1), the above mechanism of destructive interferences 
affects only the diagrams involving
the $ZZ$ fusion. On the other hand, in the diagrams involving 
$WW$ fusion (that are responsible for the dominant part of the 
basic VBF $H\,jj$ cross section)
the charged currents in the $qq^\prime W$ vertices
change the electric charges of the in-out
partons, and consequently 
the interference is now additive rather than destructive. 
Therefore, the cross section for 
$H\,\gamma\,jj$ is expected to follow the usual pattern 
of QED corrections as far as  its $WW$ fusion component 
is concerned. The relative 
contribution of the $ZZ$ fusion will be instead
remarkably smaller than in the case of the basic VBF $H\,jj$ 
process.

 In order to prove this last statement, we selected, among all 
possible
 subprocesses contributing to $pp \to H\,(\gamma) \, jj$, a 
first set of  subprocesses (named $N$)
 mediated only by the  $ZZ$ fusion, namely $q q \to H(\gamma)qq$
 [in particular, we summed up cross sections for
 $q=(u,d,s,c,\bar u, \bar d,\bar s,\bar c)$],
 and a second set of 
 subprocesses (named $C$) mediated only by the $WW$ fusion
 [in this case, we summed up the 8 cross sections of the type
  $u \bar c \to H(\gamma)\, d \bar s$].
Calling $\sigma^{(N,C)}$ the cross sections for the two latter sets,
we computed the following ratios among the radiative and the non-radiative 
processes at the LHC
\begin{eqnarray}
\frac{\sigma^{(N)}(H \gamma\, jj)}{\sigma^{(N)}(H \, jj)} = 0.0016\,,~~~~
\frac{\sigma^{(C)}(H \gamma\, jj)}{\sigma^{(C)}(H \, jj)} = 0.013\nonumber\,,
\end{eqnarray}  
where we applied the cuts $p_{\rm T}^\gamma \geq 20\, {\rm GeV}$, 
$|\eta_\gamma| \leq 2.5$,
 and  $\Delta R_{j\gamma} \geq 0.7$, assuming $m_H=120 $~GeV. 
It is then clear that the radiation is suppressed in the presence of the 
$HZZ$ vertex. 

This property enhances the sensitivity of the $H\,\gamma\,jj$ cross section
to the $WWH$ coupling. Hence, a determination of the  $WWH$ coupling
at the LHC could benefit also from a
 measurement of the $H\,\gamma\,jj$ cross section.

On the same basis, one  can 
expect that the above destructive interference effect will also
reduce the contribution from central 
photon radiation in the Higgs production through the 
$g^{\ast} g^{\ast} \to H$ process, giving also rise to 
 $H\,\gamma\,jj$ final states.
We verified that this is indeed the case. After applying the 
{\it basic} kinematical
cuts in Eq.~(\ref{eq:basic}), Section~4, with 
$p_{\rm T}^{\gamma}\geq 20$~GeV,
we obtain a reduction factor of about $8\cdot 10^{-4}$ 
for  the $H\,\gamma\,jj$
cross section arising from the 
$g^{\ast} g^{\ast} \to H$ channel with respect
to the corresponding  $H\,jj$  cross section. 
In particular, the absolute value of the $H(\to b \bar b)\gamma\,jj$ 
cross section (in the limit $m_t \to \infty$) turns out to be 0.21~fb,
which makes the impact  of the $g^{\ast} g^{\ast} \to H$ 
process on  the present analysis negligible.

%%%%%%%%%%%%%%%%%%%%
\section{Cross sections for the signal versus the irreducible background }
\label{sec4}
%%%%%%%%%%%%%%%%%%%%

The numerical results presented in this section have been independently
obtained by the Monte Carlo event generators ALPGEN~\cite{alpgen},
and MadEvent~\cite{madevent}.

The signal is calculated in the narrow width approximation, 
{\sl i.e.} we computed  the exact lowest-order matrix element for the process 
$p p \to H \gamma \, jj $, and then let the Higgs boson 
decay into a $b \bar b$ pair according to its branching ratio 
and isotropic phase space. 
After the decay,  cuts on the $b-$quark jets are implemented. 

For the irreducible $p p \to b \bar b \gamma \, jj$ background,
we computed  all the matrix elements at ${\cal O}(\alpha_s^4 \alpha) 
$, neglecting
${\cal O}(\alpha_s^2 \alpha^3)$, ${\cal O}(\alpha_s \alpha^4)$ and
${\cal O}(\alpha^5)$ contributions and their interference with the
${\cal O}(\alpha_s^4 \alpha)$ contribution. We checked that this 
has no numerical impact on the results.

The present study is limited at the parton level, apart from 
some preliminary analysis of parton-shower effects presented 
in Section~6. A more complete 
simulation, that takes into account showering, 
hadronization and detector simulation, even if 
crucial for the assessment of the potential of this channel, 
is beyond the scope of the present 
paper.  

As PDF's,  we use the parametric form of 
CTEQ5L~\cite{cteq}, and the facto\-ri\-za\-tion/re\-norma\-li\-zation 
scales are fixed at 
$\mu_{\rm F}^2 = \mu_{\rm R}^2 = \sum E_t^2$ and 
$\mu_{\rm F}^2 = \mu_{\rm R}^2 = m_H^2 + \sum E_t^2$ for the 
backgrounds and signal, respectively ($E_t$ is the transverse 
energy of any QCD parton). The three
 Higgs-mass cases  120, 130 and 140~GeV are analysed. 
 
We start by the definition of two {\it basic} event selections 
(sets 1 and 2) that differ only by the
threshold on the photon transverse momentum $p_{\rm T}^\gamma$:
\begin{eqnarray}
&&p_{\rm T}^j \geq 30\, {\rm GeV}, \, \, \, \,\, 
p_{\rm T}^b \geq 30\, {\rm GeV}, \, \, \, \,\,
\Delta R_{ik} \geq 0.7,\, \nonumber \\
&&|\eta_\gamma|\leq 2.5, \, \, \,\,\,
|\eta_b|\leq 2.5, \, \, \,\,\, |\eta_j|\leq 5, \nonumber \\
&&m_{jj} > 400\, {\rm GeV}, \, \, \, \,\,\,\,  
m_H(1-10\%) \leq m_{b \bar b} \leq m_H(1+10\%), \nonumber \\
&& 1) \, \, \, p_{\rm T}^\gamma \geq 20\, {\rm GeV}, \nonumber \\
 &&  2) \, \, \, p_{\rm T}^\gamma \geq 30\, {\rm GeV},  
\label{eq:basic}
\end{eqnarray}
where $ik$ is any pair of partons in the final state, including 
the photon, and $\Delta R_{ik} =\sqrt{\Delta^2\eta_{ik}+\Delta^2\phi_{ik}}$,
with $\eta$ the pseudorapidity and $\phi$ the azimuthal angle.
Note that the $m_{b \bar b}$ cut is only effective on the 
continuous $m_{b \bar b}$
backgrounds. On the other hand, the finite $m_{b \bar b}$ resolution 
will also affect
the following analysis of the expected signal event rate that, 
in the present
$m_H$ range,  has a  natural width for the $m_{b \bar b}$ distribution
much smaller than the $m_{b \bar b}$ resolution.
\noindent
The cross sections for the above {\it basic} event selections are 
reported 
in Table~\ref{basicxs}.
\begin{table}
\begin{center}
\begin{tabular}{||l|l|l|l|l||}\hline
& $p_{\rm T}^{\gamma, cut}$
 & $m_H = 120$~GeV  & $m_H = 130$~GeV  &  $m_H = 140$~GeV  
\\  \hline
$\sigma[H(\to b \bar b) 
\gamma jj] $ & 20~GeV& 9.3(1)~fb  & 7.4(1)~fb   & 4.74(7)~fb   \\ 
                      & 30~GeV &  6.54(7)~fb & 5.2(1)~fb     &   3.31(3)~fb
  \\  \hline
$\sigma[{b \bar b} \gamma jj] $ & 20~GeV & 406(2)~fb  
& 405(4)~fb & 389(1)~fb  \\ 
                             & 30~GeV   & 260.5(7)~fb  
& 257.9(6)~fb & 251.8(7)~fb  
\\ \hline \hline
$\sigma[H(\to b \bar b) jj] $ &  & 727(2)~fb  
& 566(2)~fb & 363(1)~fb 
\\ \hline
$\sigma[{b \bar b} jj] $ &  & 593.7(5)~pb  & 550.5(5)~pb &  505.6(4)~pb
\\ \hline
\end{tabular}            
\caption{\label{basicxs} Cross sections for  the signal 
and the irreducible
background for the {\it basic} event selections 
in Eq.~(\ref{eq:basic}). 
Higgs production cross sections include the Higgs branching 
ratios to $b\bar b$. The signal and irreducible background 
production rates for the VBF process without photon are also shown.}
\end{center}
\end{table}
Cross sections for the signal and irreducible background for the 
VBF $\,Hjj$ process (named  from now on  
``signal without photon" and ``background without photon", 
respectively) are also shown, assuming the same kinematical cuts. 

For comparison, 
requiring  the leptonic (both  $\mu^\pm$ and 
$e^\pm$) signature for the
$W$ in the $H\, W\, jj$ process~\cite{rainwater}, 
by applying the same event selection of Eq.~(\ref{eq:basic}) 
(with the constraints on the photon applied to the charged lepton, 
and $p_{\rm T}^\ell>20$~GeV), we find a 
$H(\to b\bar b)\, \ell\nu\, jj$ cross section of 2.1~fb, for 
$m_H = 120$~GeV. 
From  Table~\ref{basicxs}, one then obtains that the
 rate for the $H\, \gamma\, jj$
signal is about 4.4 times the 
$H\, \ell\nu\, jj$ rate, for $m_H= 120$~GeV. 

With the same event selection of Eq.~(\ref{eq:basic}) (Set 1), 
we also computed the cross section  for the $b \bar{b}\,\gamma\, jj$ 
final states 
mediated by an on-shell $Z$ decay  $Z \to b \bar{b}$. 
For the purely electroweak process, we obtained a cross section of 10~fb, 
while, for the QCD mediated channel,  we 
obtained a cross section of 26~fb.  As we previously noted, 
we do not expect any contamination from this background, 
 since its  typical $b \bar b$ 
invariant mass  is well below the Higgs boson mass range considered here, 
for  the foreseen experimental resolution on $m_{b \bar b}$. 

\begin{figure}[tbp]
\begin{center}
\epsfig{file=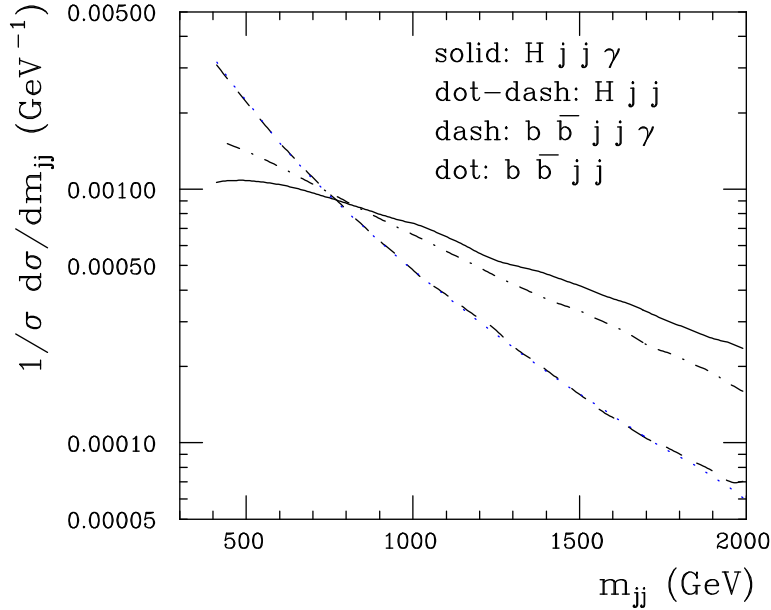,width=10cm,height=8cm, angle=0}
\end{center}
\caption{\small The two forward-jet invariant mass distribution
(critical to increase $S/B$). 
 Solid line: signal with photon. 
Dashed line: irreducible background with photon. Dot-dashed line: 
signal without photon. Dotted line: 
irreducible background without photon.
Cuts in 
Eq.~(\ref{eq:basic}) are applied, and $m_H=120$~GeV.}
\label{dmjj}
\end{figure}

\begin{figure}[tpb]
\begin{center}
\dofourfigs{3.1in}{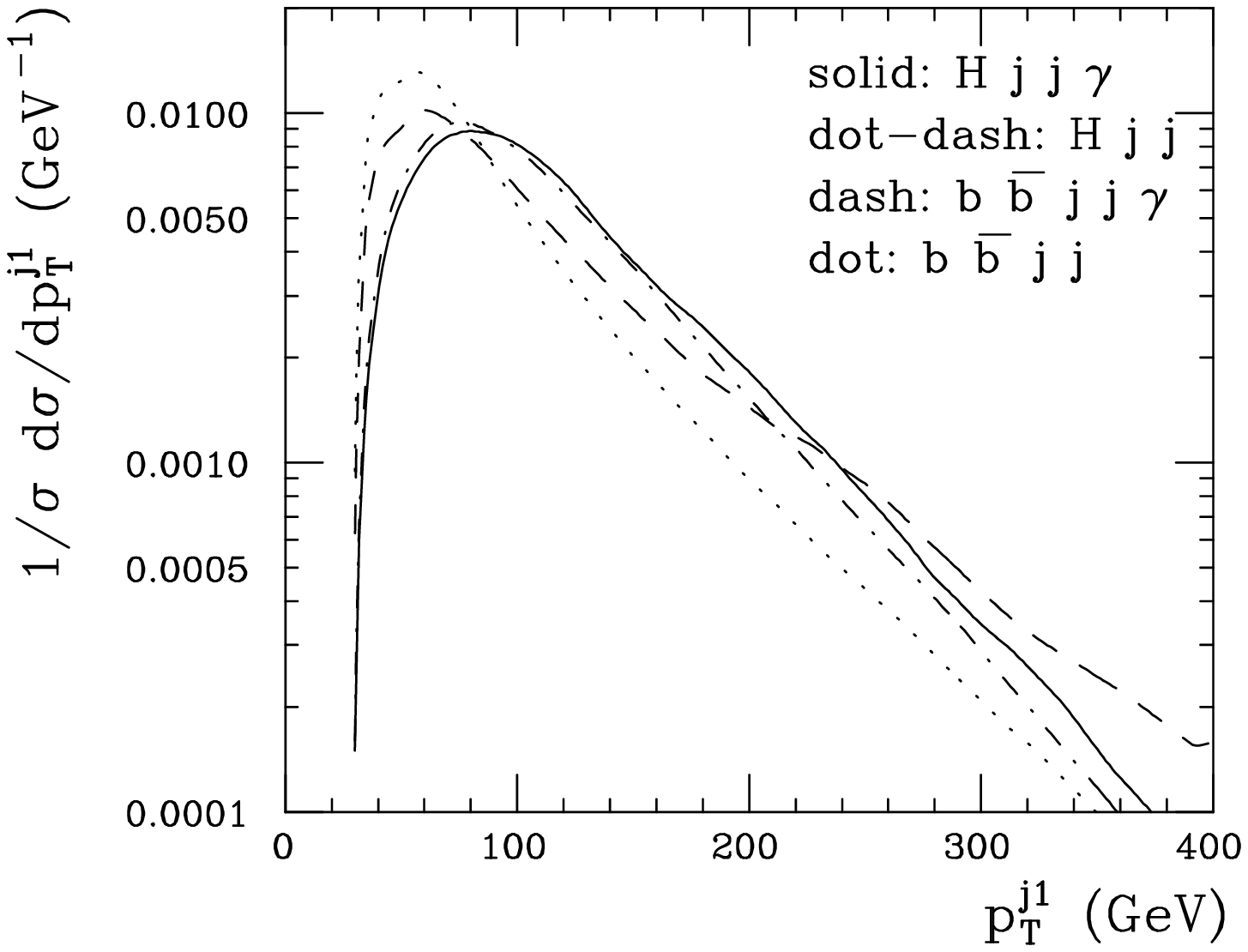}{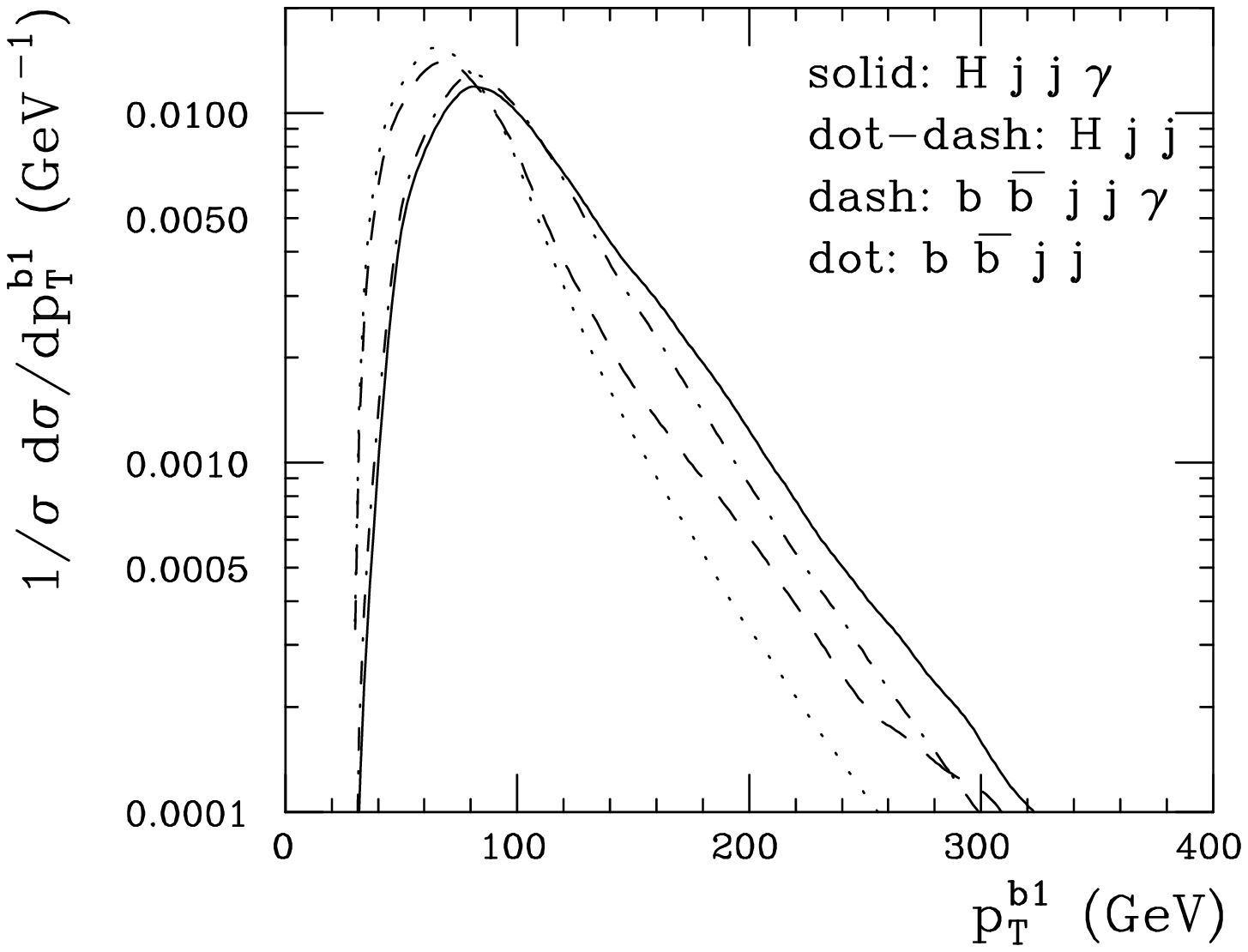}{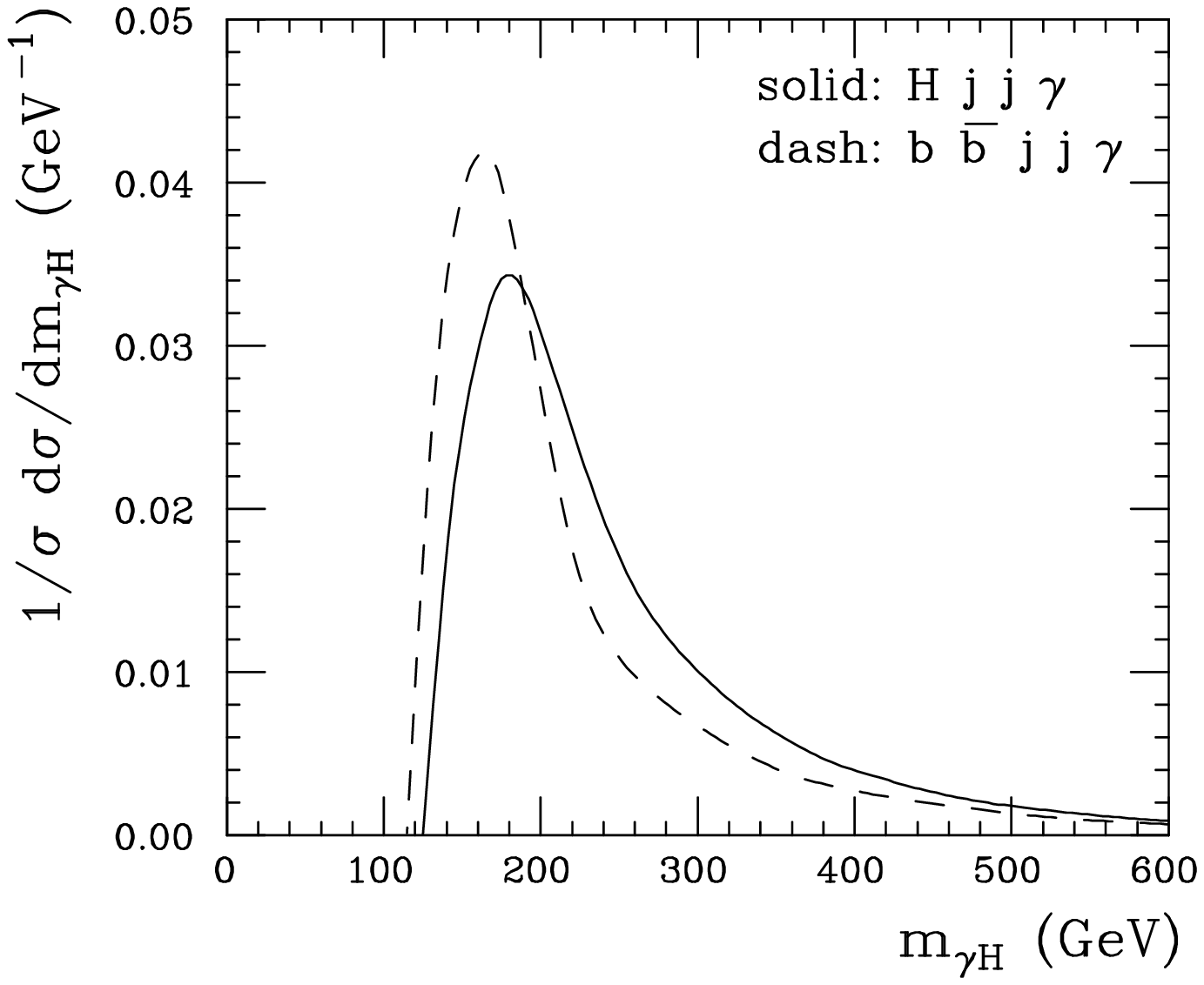}{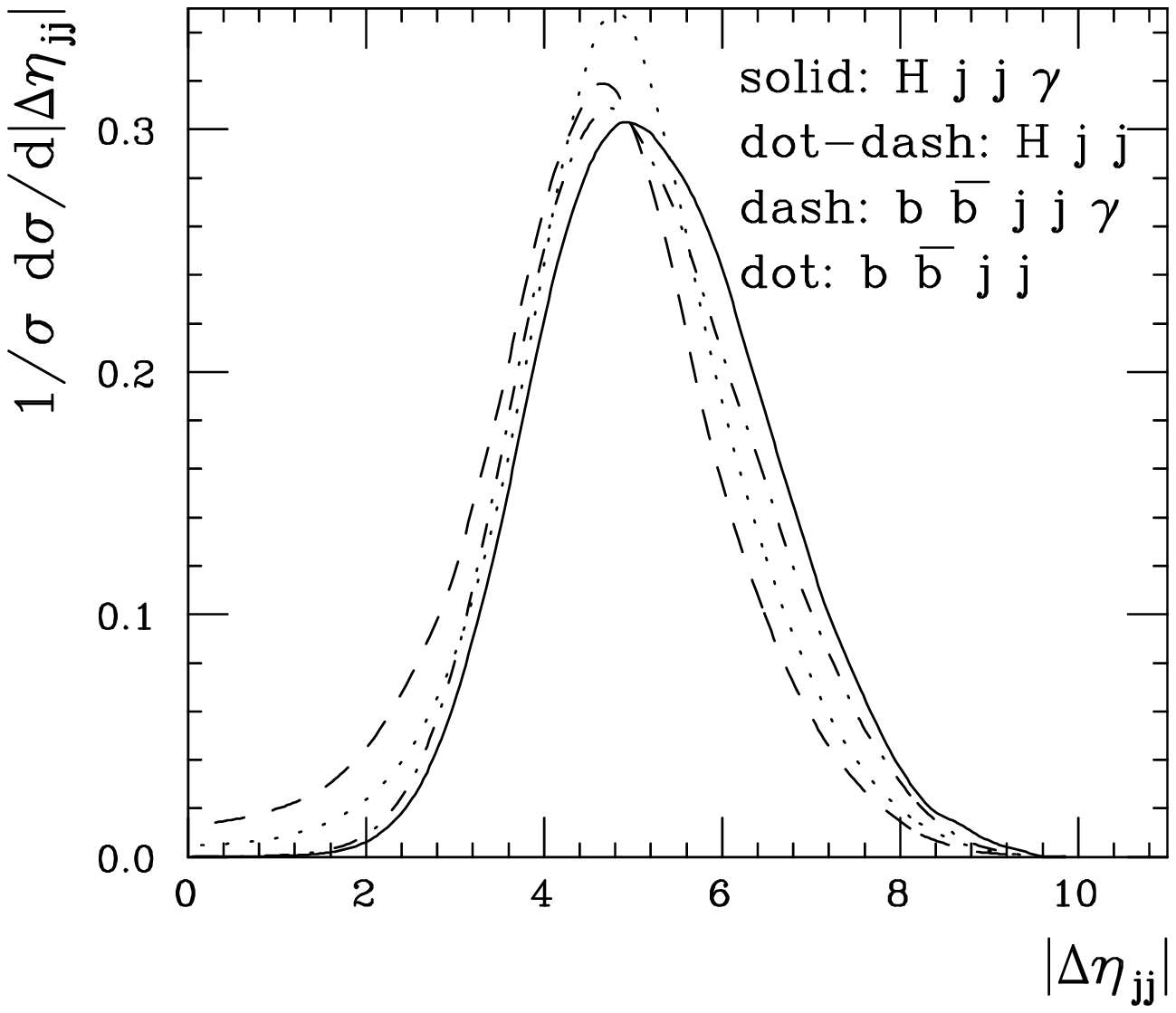}
\end{center}
\caption{\small Four further different distributions used to optimize cuts 
to  improve the $S/B$ ratio.  
Upper left panel: leading light jet $p_{\rm T}$ distribution.
Upper right panel: leading $b$ jet $p_{\rm T}$ distribution. 
Lower left panel:  invariant mass distribution for the 
Higgs boson-plus-photon system.
Lower right panel: distribution for the difference in pseudorapidity 
of the two light jets. 
Solid line: signal with photon. 
Dashed line: irreducible background with photon. Dot-dashed line: 
signal without photon. Dotted line: irreducible background without photon.
Cuts in Eq.~(\ref{eq:basic}) are applied, and $m_H=120$~GeV.}
\label{dBRx_Pe}
\end{figure}

Before comparing the signal and the background for the 
$H\, \gamma\, jj$ process, we tried to optimize our event 
selection in Eq.~(\ref{eq:basic}). Indeed,
the signal detectability can be further improved by 
imposing {\it optimized} 
cuts, that can be deduced by looking at the 
following kinematical distributions
(Figures~\ref{dmjj}, \ref{dBRx_Pe}): 
\bea
\frac{d\sigma}{d m_{jj}}, \, \, \, \, \, 
\frac{d\sigma}{d p_{\rm T}^{j1}}, \, \, \, \, \, 
\frac{d\sigma}{d p_{\rm T}^{b1}}, \, \, \, \, \, 
\frac{d\sigma}{d m_{\gamma H}}, \, \, \, \, \, 
\frac{d\sigma}{|\Delta \eta_{jj}|},   \nonumber
\eea
where $j1$ and $b1$ denote the leading ${p_{\rm T}}$ light jet 
and $b-$ jet, respectively, and $m_{\gamma H}$ is the invariant 
mass of the $\gamma b\bar b$ system.
The most effective cut in reducing the background 
with respect to the signal is related to the $d\sigma/d m_{jj}$ distribution in
Figure~\ref{dmjj}, 
as already noted in \cite{higgsplb} for the case without photon. 
 
An interesting general feature is that the shapes of 
the distributions with and without the photon are  similar,
as shown by the closeness
of the solid and dot-dashed lines on the one side, 
and of the dashed and dotted lines on the other side, in Figures~\ref{dmjj} 
and~\ref{dBRx_Pe}. 
This shows 
that the effect of QED radiation is dominated by the factorized 
eikonal approximation, 
since most of the $p_{\rm T}^\gamma$ values considered are soft 
with respect to 
the energy scale of the process. 
On the other hand, some differences between the radiative and the 
non-radiative 
processes are present, that can help significantly 
in improving the $S/B$ ratio. 

\noindent
Indeed, the request of a further central photon tends to enhance 
the characteristic kinematical features of a typical VBF event.
First of all, the $jj$ invariant mass distribution is   flatter for the 
radiative signal  than for the non radiative 
case (see solid and dot-dashed lines in Figure~\ref{dmjj}), 
while the corresponding backgrounds are almost superimposed. 
Then, increasing  the lower $m_{jj}$ cut does not imply a
dramatic reduction of the signal cross section, while it gains a substantial 
decrease of the background. Second, by adding the
photon, the distribution $d\sigma / d |\Delta \eta_{jj}|$ 
is slightly shifted toward larger $|\Delta \eta_{jj}|$ values for the signal,
while it moves in the opposite direction
for the background. 
This feature makes a cut on the pseudorapidity 
separation between the tagging jets even more effective than in the 
case of  the VBF $H\,jj$ typical event~\cite{higgsplb}. 

By studying the variation of the significance $S/\sqrt{B}$ as a function 
of the cuts on the distributions~\footnote{This was performed 
 for $m_H = 120$~GeV. For higher Higgs masses the constraints in
Eq.~(\ref{eq:optimized})  (especially the one on $m_{\gamma H}$) 
could be further tuned.}, we found an {\it optimized} event 
selection where, in addition to the {\it basic} cuts, we impose the 
following cuts~\footnote{There are few more distributions which show 
differences between signal and backgrounds. However 
these are not exploited here, since, at the parton level of the 
present analysis, they lead to an improvement in $S / B$, 
but not in $S / \sqrt{B}$.}: 
\begin{eqnarray}
&&m_{jj} \geq 800\, {\rm GeV}, \, \, \, \, \, 
p_{\rm T}^{j1} \geq 60\, {\rm GeV},  \, \, \, \, \, 
p_{\rm T}^{b1} \geq 60\, {\rm GeV}, \, \, \, \nonumber \\
&&|\Delta \eta_{jj}| > 4, \, \, \, \, \, 
m_{\gamma H} \geq 160\, {\rm GeV},  \, \, \, \, \, \Delta 
R_{\gamma b/\gamma j} \geq 1.2\, .
\label{eq:optimized}
\end{eqnarray}
With the above additional requirements, we find the cross 
sections reported in Table~\ref{optimizedxs}, 
where also the signal and irreducible background 
production rates for the VBF process without photon are  shown. 

A comment on the sensitivity to the choice of the 
factorization/renormalization scale is in order. While the 
signal is mediated by electroweak gauge bosons, and thus depends on 
$\alpha_s$ only through the PDF's, the background contains four powers 
of $\alpha_s$ and thus it is expected to be more sensitive to scale 
variations. We expect our scale choice 
$\mu_{\rm F}^2 = \mu_{\rm R}^2 = \sum E_t^2$ 
provides a conservative estimate of the background cross section, 
given the large ${\hat s}$ values of the elementary processes involved 
(arising from the large mass threshold for the pair of forward jets). We have, 
however, evaluated the impact of a variation of $\mu_{\rm F}^2$ and 
 $\mu_{\rm R}^2$ by a factor of two, obtaining a variation 
of about 25-30\% in the cross section. This translates into an uncertainty 
of the order of 10-15\% on the statistical significances studied 
in the following, thus being of moderate impact.
\begin{table}
\begin{center}
\begin{tabular}{||l|l|l|l|l||}\hline
&  $p_{\rm T}^{\gamma, cut}$ & $m_H = 120$~GeV  & $m_H = 130$~GeV  
&  $m_H = 140$~GeV  
\\  \hline
$\sigma[H(\to b \bar b)  \gamma jj] $  & 20~GeV 
&3.59(7)~fb  &2.92(4)~fb &1.98(3)~fb  \\
                                       & 30~GeV
&2.62(3)~fb  &2.10(2)~fb &1.50(3)~fb  
\\ \hline
$\sigma[{b \bar b} \gamma jj] $ & 20~GeV    &33.5(1)~fb  
&37.8(2)~fb &40.2(1)~fb 
 \\ 
                          & 30~GeV   &25.7(1)~fb  &27.7(1)~fb &28.9(2)~fb
\\ \hline \hline
$\sigma[H(\to b \bar b)  jj] $ &  & 320(1)~fb  & 254.8(6)~fb & 167.7(3)~fb  
\\  \hline
$\sigma[{b \bar b} jj] $ &  & 103.4(2)~pb  & 102.0(2)~pb & 98.4(2)~pb 
\\ \hline
\end{tabular}            
\caption{\label{optimizedxs} Cross sections for  the signal and the irreducible
background for the {\it optimized} event selections 
of Eq.~(\ref{eq:optimized}), added to the {\it basic} selection
in Eq.~(\ref{eq:basic}).  Higgs production cross sections include 
the Higgs branching ratios to $b\bar b$. The
signal and irreducible background production rates for the basic 
VBF process are also shown.
}
\end{center}
\end{table}

As anticipated in Section~\ref{sect3}, one can see in 
Table~\ref{optimizedxs} that the requirement of the extra central 
 photon with $p_{\rm T}^{\gamma}\gsim 20$~GeV in the final state 
involves a reduction factor 
of order 100 for the signal rate with respect to the final state 
without photon,  
according to the expectations of the 
 ${\cal O}(\alpha)$ QED naive scaling.
 On the other hand,
the radiative background is suppressed by a factor  of about 3000
with respect to the case of no photon radiation. As also discussed in 
Section~3, this effect can 
be understood as due to the quantum destructive interference 
 between the photon emission 
from the initial quark radiating a gluon in the $t$ channel and the 
photon emission from the corresponding final quark. 
Such an effect makes the process 
$b \bar b \gamma j j $ competitive on the statistical significance 
with the process $b \bar b j j $ studied in~\cite{higgsplb}. 

Furthermore, the presence of an additional photon can improve 
the experimental triggering efficiencies,
provided a good rejection of jets against 
photons is available~\cite{atlasnotegamma,cmsnotegamma}. 
Triggering on an inclusive 
photon with a $p_{\rm T}$ threshold of 22~GeV gives a Level-1 trigger rate 
of about 4.2 kHz in the CMS detector with an instantaneous luminosity 
${\cal L} = 2 
\times 10^{33}$~cm$^{-2}$~$s^{-1}$~\cite{cmstdr,dallavalle-sidoti}. 
Similar performances are expected for the ATLAS detector~\cite{atlastdr}. 
This rate might be further reduced with a High Level Trigger (HLT) 
based on the 
combined presence of photon plus heavy flavours or photon plus 
forward tagging jets. Such HLT studies have not yet been done. 

The effect of the inclusion  of optimized cuts as in   
Eq.~(\ref{eq:optimized}) 
 is to reduce the signal cross sections by about a factor of two, while the 
background gets scaled by about an order of magnitude, allowing to reach 
a  $S/B$ ratio for cross sections larger than 1/10  at $m_H\simeq 120$~GeV, for
both values of the $p_{\rm T}^{\gamma, cut}$. The corresponding ratio 
for the case  without photon  is about  1/300.  At  $m_H\simeq 140$~GeV, both 
the $S/B$ ratios fall down by about a factor two.

In order to evaluate the expected statistical significance of the signal,
 we assumed a $b-$tagging efficiency $\epsilon_b = 60$\%,
 and a reduction of the {\it signal} number of events by 70\% 
in the window $ m_H(1\pm10\%)$, 
due to the finite $b \bar b$ mass resolution, that broadens 
the narrow Higgs decay $b \bar b$ distribution. 
For an integrated luminosity of $100$~fb$^{-1}$, we get a 
statistical significance
 $S / \sqrt{B}|_{H\gamma\,jj}\simeq 3$, at low $m_H$ and 
$p_{\rm T}^{\gamma, cut}$
 values. At $m_H\simeq 140$~GeV and 
$p_{\rm T}^{\gamma, cut}\simeq 30$~GeV,
 it degrades down to about 1.3, mainly
 due to the falling branching ratio for $H \to b \bar b$. 
This is to be compared with $S / \sqrt{B}|_{Hjj}$  that ranges 
from 3.5 down to about 2.
 A summary of the statistical significances, including only 
the irreducible background, with an integrated luminosity of 
$100$~fb$^{-1}$ is given in Table~\ref{tab:sign1}. 
\begin{table}
\begin{center}
\begin{tabular}{||l|l|l|l|l||}\hline   &  $p_{\rm T}^{\gamma, cut}$
 & $m_H = 120$~GeV  & $m_H = 130$~GeV  &  $m_H = 140$~GeV  
\\  \hline
$S / \sqrt{B}|_{H\gamma\,jj}$ & 20~GeV & 2.6  & 2.0 & 1.3  
\\ \hline
$S / \sqrt{B}|_{H\gamma\,jj}$ & 30~GeV  & 2.2  & 1.7 & 1.2 
\\ \hline \hline
$S / \sqrt{B}|_{H\,jj}$ &  & 3.5  & 2.8 & 1.9 
\\ \hline
\end{tabular}            
\caption{\label{tab:sign1} Statistical significances 
with the event selection of Eq.~(\ref{eq:basic}) and 
(\ref{eq:optimized}), with an integrated  luminosity 
of $100$~fb$^{-1}$. The value $\epsilon_b = 60$\% for the 
$b-$tagging efficiency and a Higgs boson event reduction 
by $\epsilon_{b \bar b}\simeq$ 70\%, due to the finite ($\pm$10\%)
$b \bar b$ mass resolution,  have been assumed. Jet-tagging efficiency and 
photon-identification efficiency are set to 100\%. 
Only the irreducible background is included in this analysis.}
\end{center}
\end{table}

%%%%%%%%%%%%%%%%%%%%%%%%%%%%%%%%
\section{Reducible backgrounds}
\label{sect:reducible}
%%%%%%%%%%%%%%%%%%%%%%%%%%%%%%%%
A complete analysis of the reducible backgrounds to the   
$H\,\gamma\,jj$    signal is beyond the scope of our study. 
For instance, the potential dangerous contamination 
coming from $\pi^0$ decays into photons can only be studied with a simulation 
including showering, hadronization and detector simulation, and is left 
to a further investigation. 
However, in order to have a sensible estimate of the 
achievable $S/B$ ratio and statistical significance at parton level, 
we computed with ALPGEN the cross sections, assuming $m_H=120$~GeV and
with the optimized event selection of Eq.~(\ref{eq:basic}) 
and (\ref{eq:optimized}),  
for three main potentially dangerous processes~\footnote{
We estimated that the process $p \bar p \to c \bar c \gamma j j$, where
the $c$ quarks are both mistagged as $b$ quarks, 
(assuming $\epsilon_c=10\%$) can be safely  neglected.}: 
\begin{itemize}
\item $p p \to \gamma + 4 $~jets, where two among the light jets are 
fake tagged as $b-$jets;
\item $p p \to b {\bar b} + 3$~jets, where one of the light jets is 
misidentified as a photon;
\item $p p \to 5$~jets, where one of the light jets is misidentified as 
a photon, and two light jets are fake tagged as $b-$jets. 
\end{itemize}
With more than two light jets in the final state, the selection 
criteria need to 
be specified, in order to avoid ambiguities. Even if some algorithm able 
to mimic a realistic analysis should be implemented, we adopt a very simple 
one, which however should give us a sound estimate of the cross sections. 
At first, we look for the 
pair of light jets with the largest invariant mass, and identify the latter
as the tagging jets. For the channel $\gamma + 4$~jets, this is enough 
to guarantee also a unique assignment for the fake $b-$jets. In the case 
of the $b \bar b + 3$~jets final state, once we have specified the 
tagging jets, the remaining light jet is the one misidentified as the photon. 
An additional specification is required for the $5$~jet channel. After 
the tagging jet selection, we choose the photon at random among the remaining 
three light jets. Then,  the remaining two jets are the fake $b-$tagged jets. 
With the above specifications of the event selection, keeping the 
same setup for the energy scales and PDF's as in the previous Section, 
we obtain the cross sections quoted in Table~\ref{reducible}. 
\begin{table}
\begin{center}
\begin{tabular}{||l|l|l||}\hline
 & $p_{\rm T}^\gamma \geq 20$~GeV  & $p_{\rm T}^\gamma \geq 30$~GeV   
\\  \hline
$\sigma(p p \to \gamma + 4j)$ & 2.27(2)~pb  & 1.72(4)~pb   
\\  \hline
$\sigma(p p \to b {\bar b} + 3j)$ &61.1(3)~pb & 45.1(2)~pb  
\\  \hline
$\sigma(p p \to 5j)$ & 2.40(1)~nb & 1.83(1)~nb 
\\  \hline
\end{tabular}
\caption{\label{reducible} Cross sections for reducible 
background channels, for the optimized event selections 
of Eq.~(\ref{eq:basic}) and (\ref{eq:optimized}), 
applied as explained in the text.
 The value $m_H=120$~GeV is assumed. 
 }
\end{center}
\end{table}
The quoted cross 
sections should then be multiplied by the appropriate efficiencies: 
$\epsilon_{\rm fake}^2$ for $p p \to \gamma + 4 $~jets, 
$\epsilon_{b}^2 \epsilon_{\gamma j}$ for $p p \to b \bar b + 3$~jets and 
$3 \,\epsilon_{\rm fake}^2\, \epsilon_{\gamma j}$ for $p p \to 5$~jets, 
where $\epsilon_{\rm fake}$ is the efficiency of mistagging a  light jet
as a $b-$jet, and $\epsilon_{\gamma j}$ is the rejection 
factor of a jet against a photon. 
Adopting the same efficiencies as  in  Table~\ref{tab:sign1}, and 
assuming
$\epsilon_{\rm fake} = 1$\% and 
$\epsilon_{\gamma j} = 1/5000$\footnote{This is the value 
quoted in~\cite{atlasnotegamma}, see also~\cite{cmsnotegamma}.}, we 
obtain  the event numbers  for signal and backgrounds  shown in 
Table~\ref{evts}, assuming  $m_H=120$~GeV and an integrated 
luminosity of $100$~fb$^{-1}$. 
\begin{table}
\begin{center}
\begin{tabular}{||l|l|l||}\hline
  & $p_{\rm T}^\gamma \geq 20$~GeV  & $p_{\rm T}^\gamma \geq 30$~GeV   
\\  \hline
$p p \to \gamma H(\to b \bar b) + 2 j$      & 90  & 66   
\\  \hline
$p p \to \gamma b \bar b + 2j$              & 1206  & 925 
\\  \hline
$p p \to \gamma + 4j$                       & 23  & 17 
\\  \hline
$p p \to b {\bar b} + 3j$                   & 440 & 324   
\\  \hline
$p p \to 5j$                                & 14 & 11 
\\  \hline
$S / \sqrt{B}$                                & 2.2 & 1.8
\\  \hline
\end{tabular}
\caption{\label{evts} Event numbers for signal, irreducible 
and reducible backgrounds, for the case $m_H = 120$~GeV, 
with an integrated luminosity of $100$~fb$^{-1}$. The efficiencies 
are described in the text. The last line shows the statistical significance 
including all the background channels.}
\end{center}
\end{table}
As can be seen in the last line of Table~\ref{evts}, by including 
also the reducible backgrounds, 
the statistical significance  decreases by 
about 14(12)\% for $p_{\rm T}^{\gamma, cut}=20(30)$~GeV)
with respect to Table~\ref{tab:sign1}, where only the 
irreducible background has been considered. The most dangerous 
contribution to reducible backgrounds comes from $p p \to b \bar b + 3 j$. 

A possible way of increasing the number of signal events is to require 
at least one $b-$tag instead of two. In fact, with $\epsilon_b = 60$\%, 
the effective tagging efficiency becomes 
$2\epsilon_b(1-\epsilon_b)+\epsilon_b^2=0.84$, instead of $\epsilon_b^2=0.36$. 
Then, considering only the irreducible background,
the $S/B$ ratio remains unchanged but the significance
increases by a factor of about 1.5.  
We have also performed a preliminary analysis
of event numbers assuming  single $b-$tagging. 
Omitting potentially large reducible backgrounds, such as 
multi-jet processes with one
heavy quark and a mistagged jet, we found a significance 
slightly below the one obtained in the two $b-$tag case. 
This conclusion, however, relies strongly on the assumed 
fake tag probabilities, which can be estimated only with a 
full detector simulation. It could therefore be worthwhile to 
investigate also this possibility. 

\section{Parton-shower effects and central jet veto}
A feature of the signal is that to leading order no colour is exchanged 
between the up and down fermionic lines of Figure~1, since 
$p p \to H \gamma \, jj $ is an electroweak mediated 
process~\cite{rapiditygap}. Thus the typical scale for QCD radiation 
is $p_{\rm T}^j$. 
On the contrary, the diagrams of the 
background processes are characterized by the presence 
of $t-$channel virtual gluons (cf. Figure~4). Thus the 
typical scale for QCD radiation is of the order of 1~TeV ($m_{jj} + m_H$). 
On these grounds, we can expect that higher-order QCD radiation 
will be much more relevant for the background than for the signal. 
The fairly different radiation pattern can be exploited to further 
enhance the $S/B$ ratio, as suggested in \cite{rapiditygap}. 
We investigated qualitatively these features 
by simulating higher-order QCD radiation with the HERWIG 
parton shower~\cite{herwig} on top 
of the partonic unweighted events, generated with ALPGEN with the 
optimized event selection of Eqs.~(\ref{eq:basic}) and (\ref{eq:optimized}), 
set~1. This is not a completely consistent 
approach, since one is sensitive to the partonic event selection, which 
has no physical meaning. A more solid study could be done by using 
a consistent matching procedure 
between parton shower and multi-parton matrix 
elements~\cite{ckkw}, which amounts to merging together event samples 
originating from different partonic multiplicities. We have however 
verified that, restricting our analysis to a sample with cuts tighter 
than parton level cuts, results are essentially unchanged, and we are 
therefore confident that a more refined study would lead to not too 
different results. 

Since the aim of this section is simply to give 
a qualitative estimate of the effects of a realistic event simulation, 
we will restrict ourselves to the naive procedure of showering the events 
with partonic cuts corresponding to the final event selection. 
The value $m_H=120$~GeV is considered. 
\begin{figure}[tpb]
\begin{center}
\dosixfigs{3.1in}{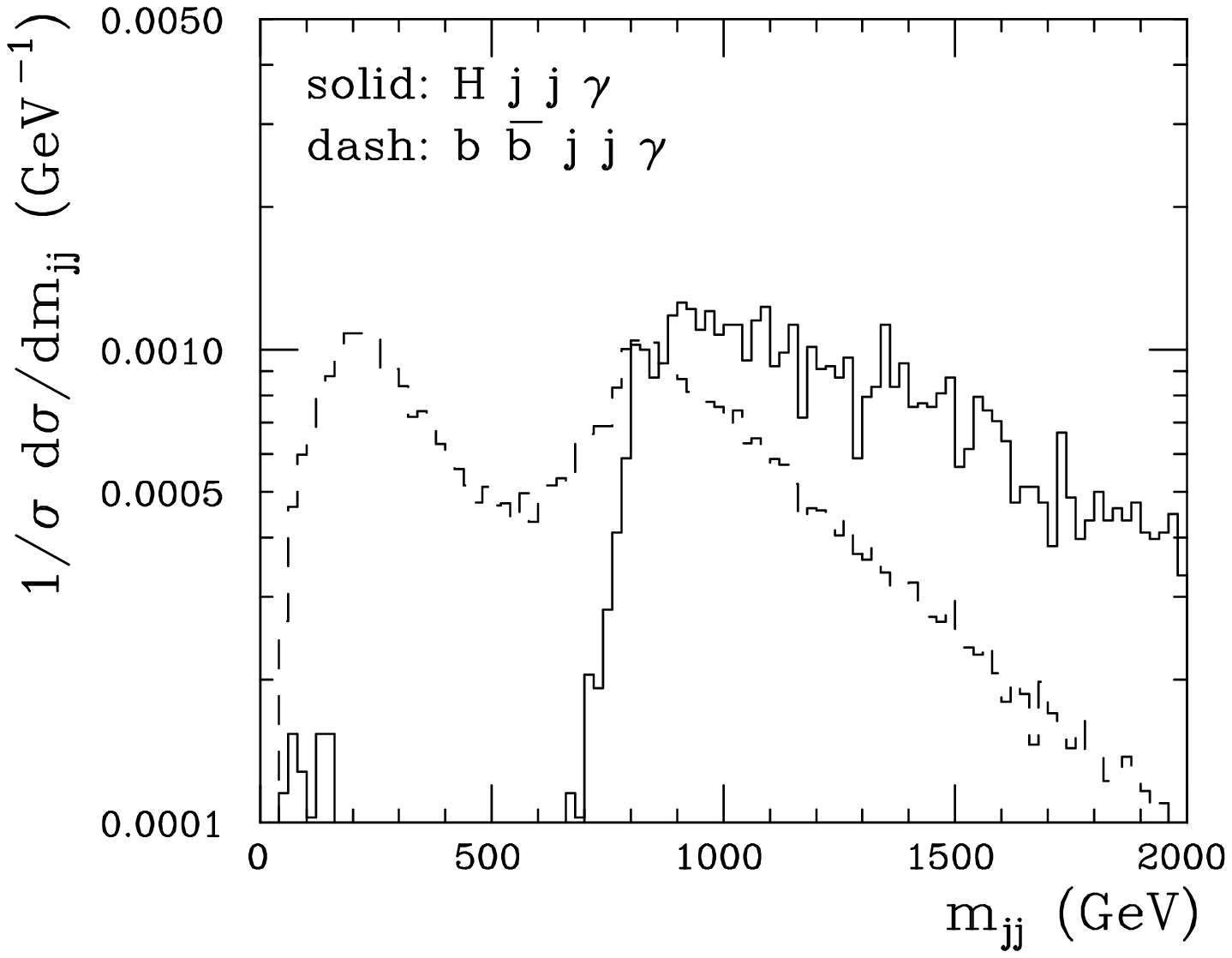}{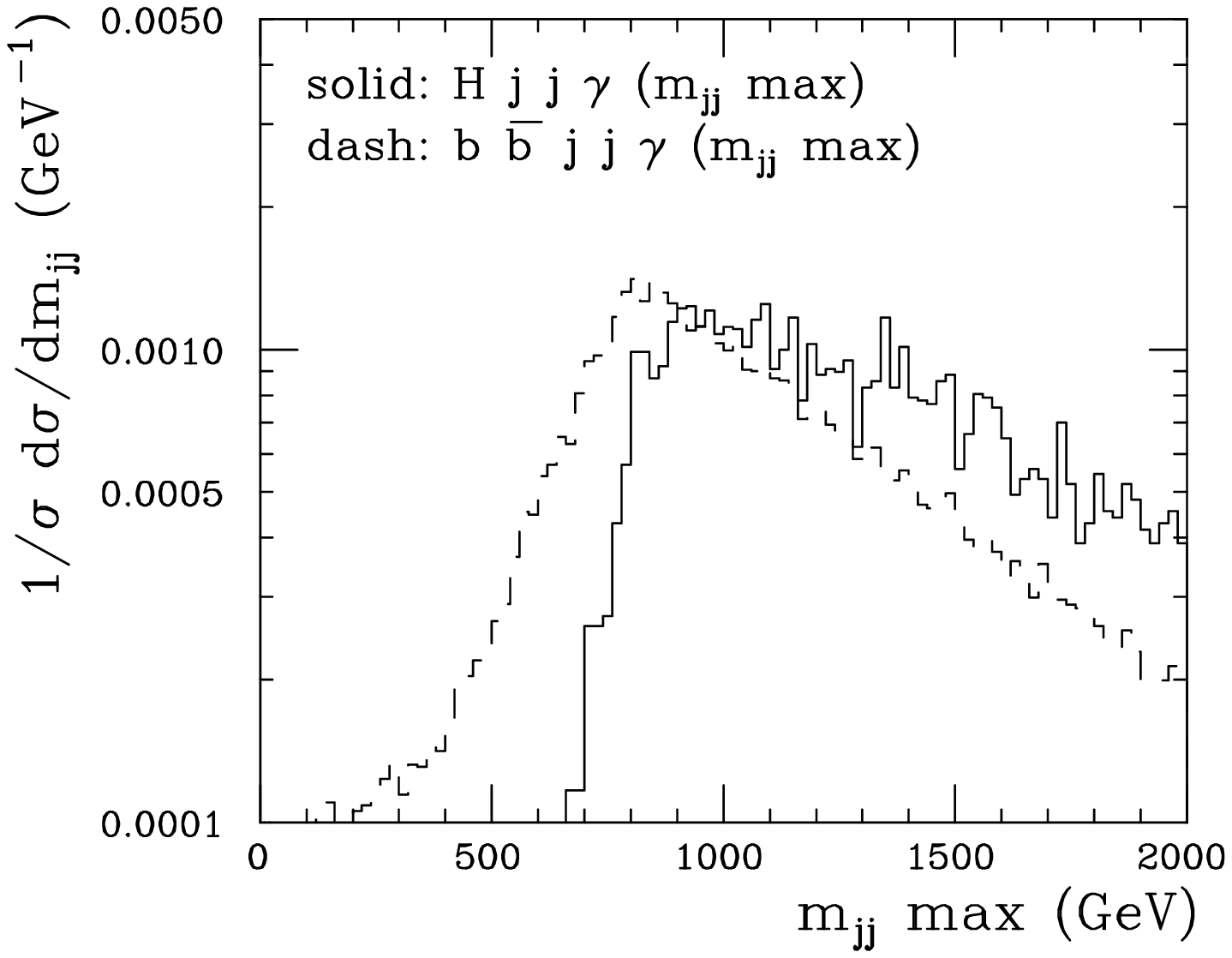}{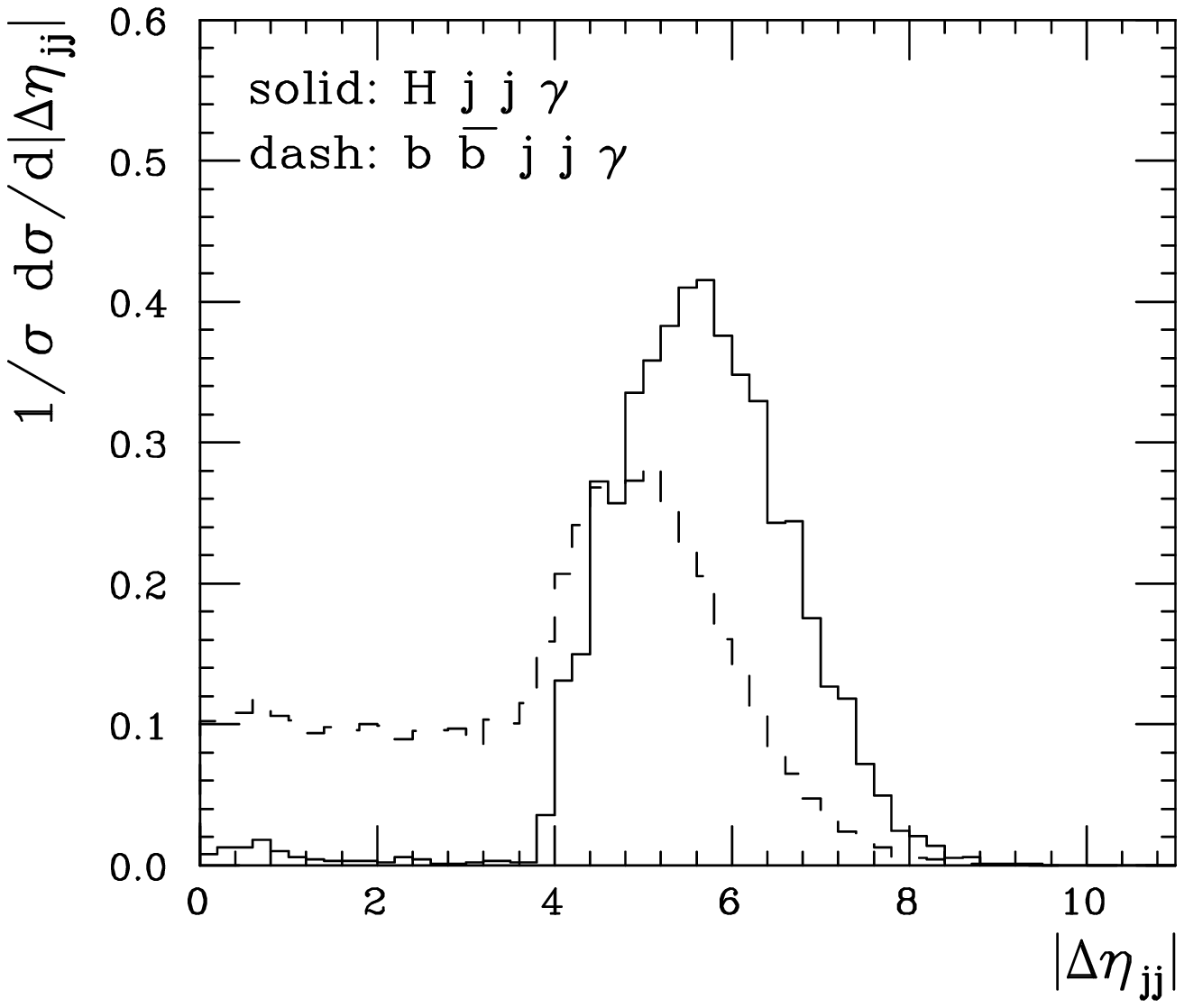}{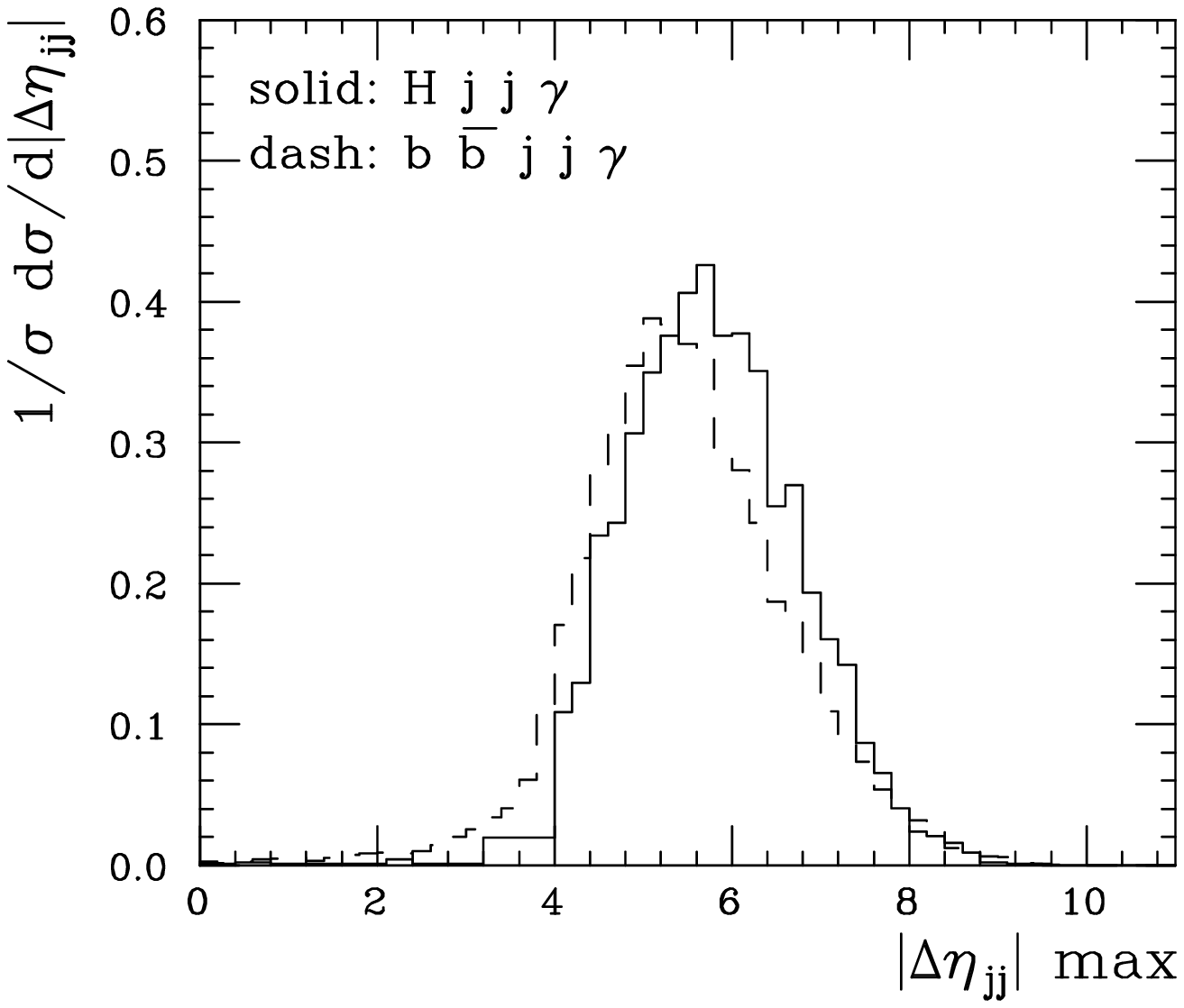}{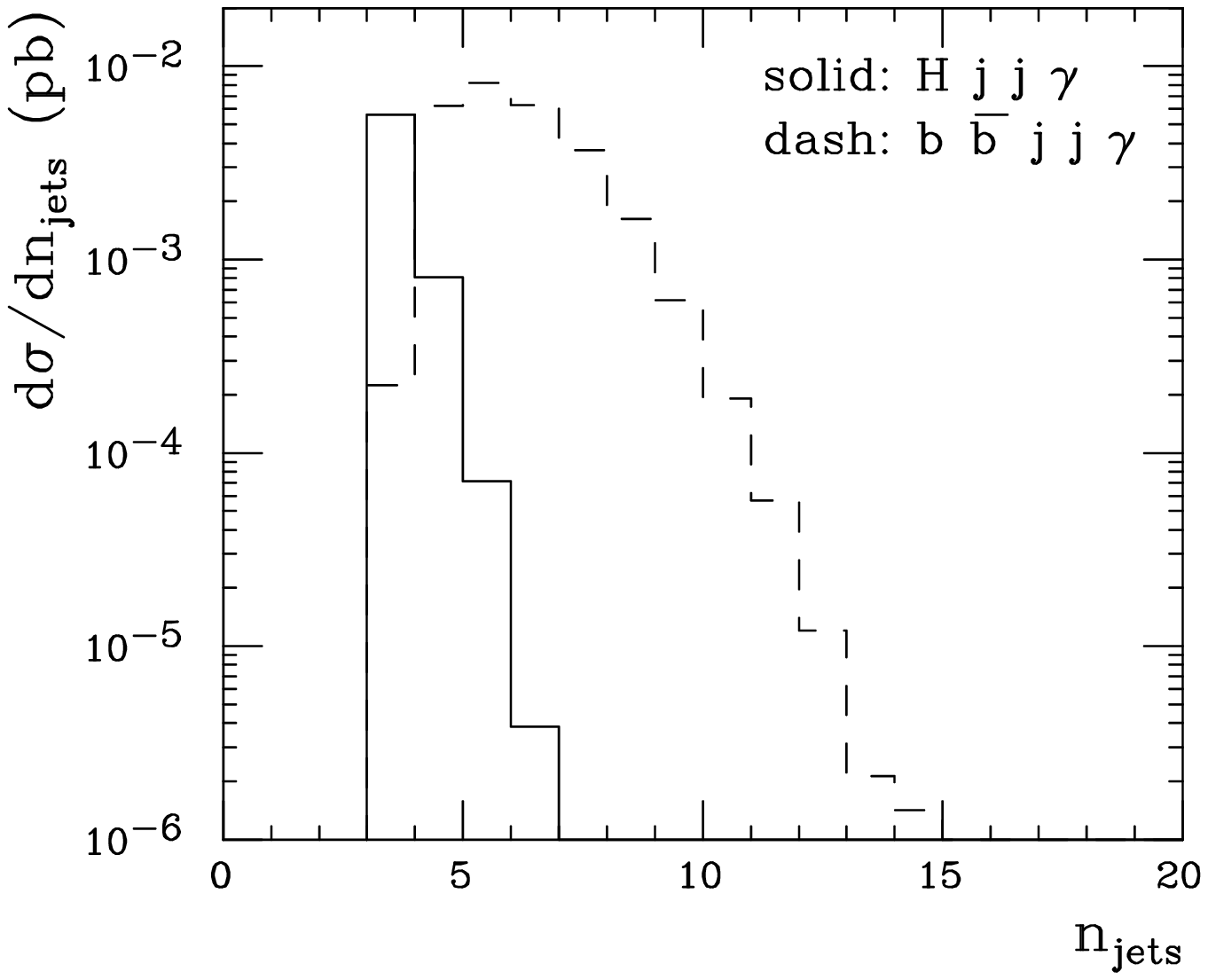}{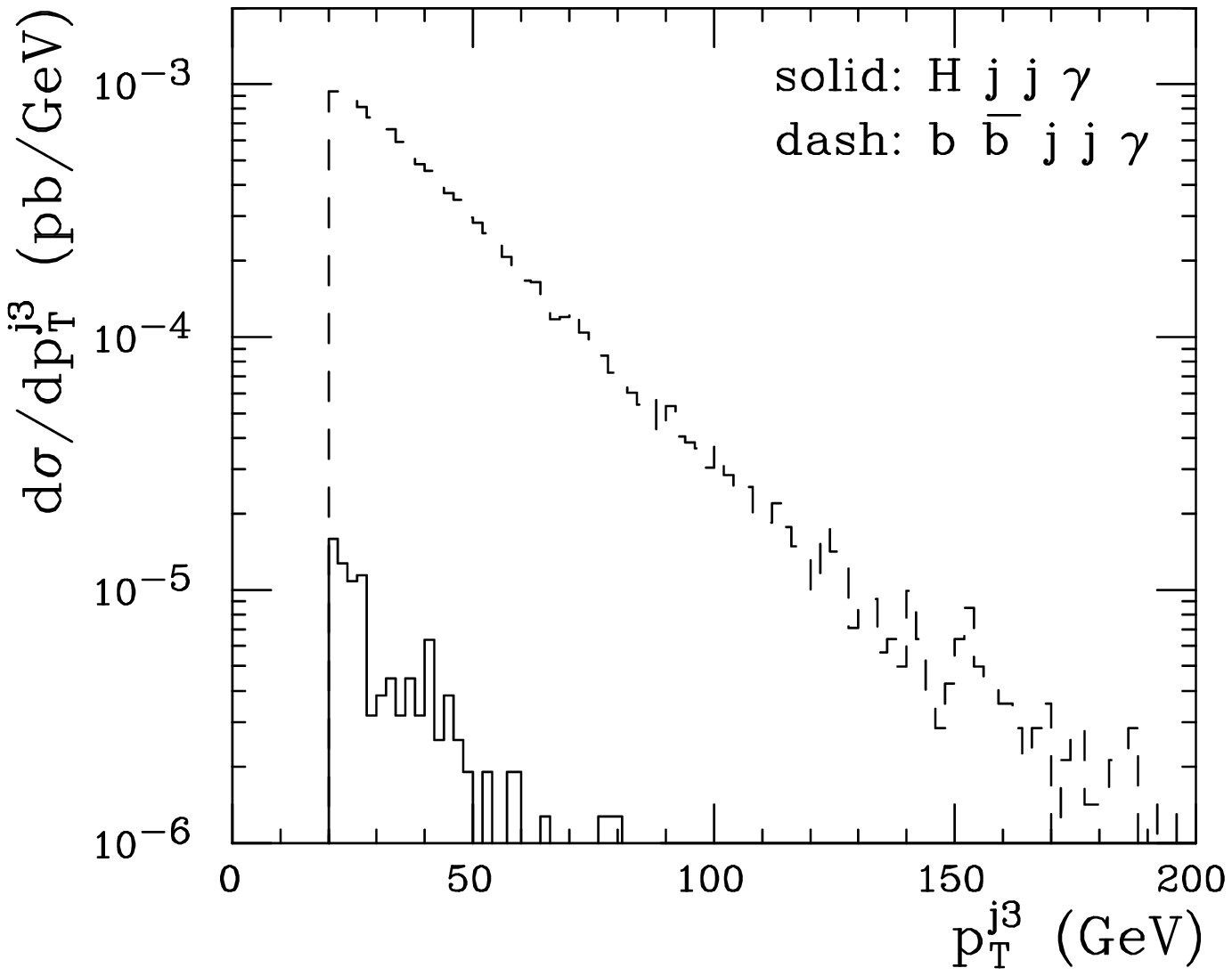}
\label{dshower}
\end{center}
\caption{\small 
First row: $m_{jj}$ distribution of the highest and second highest 
$p_{\rm T}$ jets after shower for signal and background (left), and 
maximum $m_{jj}$ distribution among all possible jet pairs satisfying 
the tagging requirements (right). 
Second raw: $ |\Delta \eta_{jj} |$ distribution for 
the highest and second highest $p_{\rm T}$ jets after shower (left), 
and maximum pseudorapidity separation between jets satisfying the 
tagging $p_{\rm T}$ thresholds (right). 
Third row: jet multiplicity distribution for signal and background (left); 
$p_{\rm T}$ distribution of the third highest $p_{\rm T}$ jet (right).}
\end{figure}
Jets are defined via 
a cone algorithm using the routine GETJET~\cite{getjet}, which uses a 
simplified version of the UA1 jet algorithm, with parameters given by 
\begin{equation}
p_{\rm T}^j > 20\,, {\rm GeV}\,, \, \, \, \, \, \, \, 
\vert \eta_j \vert < 5\,, \, \, \, \, \, \, \, 
R=0.7\,, 
\end{equation}
where $R$ is the jet cone radius. 
The $b-$tagged jets are defined as the ones containing the 
original $b$ quarks. 

Given the presence of extra QCD radiation, 
the identification of light tagging jets, among the remaining ones, 
is not uniquely defined. 
We explored two different algorithms: as a first choice, 
the tagging jets are identified 
by the highest and second highest $p_{\rm T}$ jets (referred to as 
algorithm $a1$); an alternative is to identify the tagging jets 
as the pair of jets with the highest invariant mass (algorithm $a2$). 
For both event selections we require $p_{\rm T}^{j1} \geq 60$~GeV and 
$p_{\rm T}^{j2} \geq 30$~GeV. The results are shown in 
Figure~8. In particular, the panels in the 
first raw show the invariant mass distribution of signal and background 
(solid and dashed histogram, respectively), 
for tagging jet identifications $a1$ and $a2$. 
While the signal is practically insensitive 
to the choice between the algorithm $a1$ and $a2$, the background 
shows a large difference. With the choice $a1$, a large fraction of events 
gives rise to a sort of peak at low invariant masses. 
Using this choice to select the tagging jets, 
while a cut on the invariant mass of the tagging jets 
$m_{jj} \geq 800$~GeV, applied after the shower, would reduce the signal 
only by about 7\%, the same cut would reduce the background 
by a factor of about two. 
We expect the peak at low invariant masses to be due to the identification 
of tagging jets with jets originated from the shower and not containing 
the original hard partons. This interpretation is confirmed by the 
panels in the second row, 
which shows the $\vert \Delta \eta_{jj} \vert$ distributions 
for signal and background (solid and dashed histogram, respectively). 
On the left the eta separation between the tagging jets, 
defined according to algorithm $a1$, is shown. While for almost all 
signal events one has $\vert \Delta \eta_{jj} \vert \geq 4$, a 
consistent portion of background events displays a smaller rapidity 
separation. In these events, at least one of the tagging jets 
originates from higher order parton shower emission. 
By looking for the pair of jets satifying the tagging requirement 
$p_{\rm T}^{j1} \geq 60$~GeV and $p_{\rm T}^{j2} \geq 30$~GeV, with 
the additional constraint of maximum pseudorapidity separation, 
the correspondence between tagging jets and parton level jets 
is restored, as shown in the panel on the right. 

The lower panels of Figure~8 give an estimate 
of the jet multiplicity 
(left) in signal (solid line) and background (dashed line), 
after $a1$ event selection has been applied with the additional cut 
on $m_{jj}$ at the jet level. In this case we increased the threshold 
up to $1000$~GeV, in order to minimize possible dependence 
on parton level cuts. 
While in the former case the jet multiplicity is sharply peaked 
at the value of four, in the latter case a much broader spectrum 
is present. On the right panel the 
$p_{\rm T}$ distribution of an additional $p_{\rm T}$ ordered jet with 
$p_{\rm T} \geq 20$~GeV and pseudorapidity between the tagging jet 
$\eta$'s is shown. If one imposes a further cut $m_{jj} \geq 1000$~GeV 
(the result is only slightly changed if $m_{jj} \geq 800$~GeV is used), 
a fraction of the order of 60\%(50\%) 
of the background events contains at least 
an extra jet with $p_{\rm T} \geq 20(30)$~GeV, while only a small fraction 
(of the order of 2\%) of the signal events contain such additional radiation. 
Thus a veto on additional jet activity in the central rapidity region, 
proposed for heavy and light Higgs searches in \cite{jetveto},  
could be very effective in suppressing the background more than the signal. 

In principle, an additional handle to further 
suppress the background is given by the 
photon isolation criteria. For instance, we checked that, 
with our simplified calorimeter, 
a good choice could be to require no charged tracks with 
$p_{\rm T} \geq 1$~GeV within a cone $\Delta R \simeq 0.2$ around the photon. 
For a quantitative 
statement, a full detector simulation is required. However, 
we expect that, for the present event selection, this kind of 
photon isolation criteria would actually lead to a small improvement 
in significance, rather than lowering the significance, 
as one would naively expect.  

Even if a more refined analysis will be necessary for quantitative statements, 
we expect that the background events could be lowered by a factor of about 
four with respect to the partonic estimates of previous sections, once 
showering effects are included in the analysis and  
a central jet veto strategy is adopted. 
On the other hand, the signal seems to be 
almost insensitive to shower and central jet veto effects. This would lead to 
an improvement of about a factor of two in the signal significance. 

\section{Conclusions}
In this paper, we studied the detectability of the Higgs boson 
production signal,
when the Higgs boson is accompanied by a high$-p_{\rm T}$ 
central photon and two forward jets at the LHC. The Higgs boson 
decay into a $b \bar b$ pair is considered. We analyzed the signal, 
the irreducible QCD background, 
and main reducible backgrounds at the parton level.
 The presence of a photon in the final state can improve 
the triggering efficiencies
 with respect to the basic VBF Higgs production without a photon. 
Moreover, we find that the requirement of a central 
photon in addition to the typical VBF final-state topology 
significantly suppresses the irreducible 
 QCD background. In particular, the latter has rates that are lower than
 the expectations of the 
 ${\cal O}(\alpha)$ QED naive scaling by more than an order of magnitude.
 We discussed thoroughly  the dynamical effects that are responsible for this 
 reduction. 
 As a consequence, after optimizing kinematical cuts, we obtain a 
statistical significance $S/\sqrt{B}$ for the 
 $H(\to b\bar b)\gamma\,jj$   channel  that goes from around 3, 
if  $m_H\simeq 120$~ GeV, down  to
about  1.5,  if $m_H\simeq 140$~ GeV,
for an integrated  luminosity of $100$~fb$^{-1}$. 
These significances are not far from the corresponding values 
for  the basic $H(\to b\bar b)\,jj$ process without a photon. 
The latter estimates
are based on the irreducible QCD background. 
The impact of including a few main reducible backgrounds has 
also been studied, and found to be moderate.

\noindent
A preliminary analysis of parton-shower effects points to a  
further differentiation between the signal and background 
final-state topology and composition. In particular, a preliminary 
analysis of showering and central jet veto effects points to 
an improvement of $S / \sqrt{B}$ by a factor of two. 

The same dynamical effects that are responsible for the 
irreducible background suppression also remarkably curb 
the relative contribution of the 
$ZZ\to H$ boson fusion 
diagrams with respect to the $WW\to H$ ones  
in the process $pp\to H(\to b\bar b)\gamma\,jj$.
As a consequence, we think that the study of the 
$H(\to b\bar b)\gamma\,jj$  signal at the LHC could 
have a role in the determination of both the $Hbb$ and $HWW$ couplings.
Further studies, including complete showering, hadronization, 
and detector simulations, that are beyond the scopes of the 
present paper, will be needed to establish the actual 
potential of the process $H(\to b\bar b)\gamma\,jj$ in this field.
 
\section*{Acknowledgments}
We  wish to thank  Anindya Datta for his  contribution during 
the early stages of this work, and Paul Hoyer and Antonello Polosa 
for useful discussions. We thank Alexandre Nikitenko and Antonio Sidoti for 
many useful discussions on trigger issues. 
E.G., M.M. and F.P. also wish to thank the 
CERN Theory Unit for partial support and hospitality. 
B.M.'s and R.P.'s research was partially supported by the 
RTN European Programme MRTN-CT-2006-035505
(HEPTOOLS, Tools and Precision Calculations for Physics Discoveries at
Colliders). 
B.M.'s research was also partially supported by the RTN European Programme 
MRTN-CT-2004-503369 (Quest for Unification). The work of E.G. was supported 
by the Academy of Finland (project number 104368). 
The research of R.P. was supported in part by 
MIUR under contract 2006020509\_004.

\end{document}